\documentclass[authoryear,12pt]{elsarticle} 
\usepackage[english]{babel}
\usepackage{amsmath}
\usepackage{subcaption}
\usepackage[toc,page]{appendix}
\usepackage{amssymb}       
\usepackage{amsthm}       
\usepackage{graphicx}
\usepackage{booktabs}
\usepackage{color}
\usepackage{gensymb} 
\usepackage{tabularx} 

\journal{European Journal of Mechanics A/Solids, accepted for publication}

\newcommand{\tensor}[1]{\boldsymbol{#1}}
\newcommand{\ftensor}[1]{\mathbb{#1}}

\newcommand{\rd}{\mathrm{d}}

\begin{document}

\begin{frontmatter}

\title{A generalized line tension model for precipitate strengthening in metallic alloys} 

\author{R. Santos-G{\"u}emes$^{1, 2}$}
\author{J. Segurado$^{1, 2}$}
\author{J. LLorca$^{1, 2, }$\corref{cor1}}
\address{$^1$ IMDEA Materials Institute, C/ Eric Kandel 2, 28906, Getafe, Madrid, Spain. \\  $^2$ Department of Materials Science, Polytechnic University of Madrid/Universidad Polit\'ecnica de Madrid, E. T. S. de Ingenieros de Caminos. 28040 - Madrid, Spain.}

\cortext[cor1]{Corresponding author}

\begin{abstract}

A generalized line tension model has been developed to estimate the critical resolved shear stress in precipitation hardened alloys.  The model is based in previous line tension models for regular arrays of either impenetrable or shearable spherical precipitates that were expanded to take into account the effect of the elastic mismatch between the matrix and the precipitates. The model parameters are calibrated from dislocation dynamics simulations that covered a wide range of precipitate diameters and spacing as well as of the mismatch in elastic constants. This model is  extended to deal with random arrays of monodisperse spherical precipitates by changing the geometrical parameters of the model by the averaged ones corresponding to the random distributions. The model predictions were in good agreement with the critical resolved shear stresses obtained from dislocation dynamics simulations of random spherical precipitates distributions for both impenetrable and shearable precipitates, providing a fast and accurate tool to predict  precipitate strengthening in metallic alloys. 
\end{abstract}


\end{frontmatter}

\section{Introduction}
\label{Sec:Intro}

Precipitation hardening is one of the most effective strategies to increase the yield strength of metals, and it is always present in high strength alloys \citep{Ardell1985,Kelly1971,Martin1998}. Precipitation hardening is achieved by the dispersion in the metallic matrix of second phases or precipitates (with average size in the range of a few nm to 1 $\mu$m) that hinder dislocation motion. The efficiency of precipitation hardening is known to depend on geometrical factors (size, shape, volume fraction and spatial distribution of the precipitates) and on the actual mechanisms of dislocation/precipitate interactions (either dislocation looping or shearing), which are also affected by the differences in elastic constants between the matrix and the precipitate, misfit strains, etc. \citep{Nembach1997}.

Due to the importance of this mechanism, modeling of precipitation hardening has been a very active area of research by means of analytical or numerical approaches. The latter generally involve complex simulations based on molecular mechanics for the case of very small precipitates ($<$ 10 nm) \citep{Singh2010,Bonny2011,Saroukhani2016,Esteban-Manzanares2019b,Esteban-Manzanares2019c,Esteban-Manzanares2019a} or dislocation dynamics (DD) for larger ones \citep{Xiang2006,Mohles1999,Mohles2001a,Monnet2011,Queyreau2010}. DD simulations have become more accurate in recent years, as the most relevant mechanisms controlling dislocation/precipitate interactions  (elastic mismatch, coherency strains, actual shape and spatial distribution of the precipitates, etc.) were included in the simulation framework \citep{Takahashi2008, Takahashi2011, Lehtinen2016, Bocchini2018, Santos-Guemes2018, Santos-Guemes2020, Santos-Guemes2021, Hu2021}.

Nevertheless, numerical predictions of the critical resolved shear stress (CRSS) to overcome the precipitates using either molecular mechanics or dislocation dynamics require a large computational effort. It would be very convenient to have analytical expressions that can provide fast and accurate estimations of the CRSS for actual metallic alloys also taking into account the most relevant mechanisms of dislocation/precipitate interactions. This type of approaches were pioneered by Orowan \cite{Orowan1948}, who developed a line tension model to study precipitation hardening in the case of  a square array of impenetrable circular precipitates on the glide plane. The CRSS was expressed as 

\begin{equation}
\tau_c^{Orowan}=\beta\frac{Gb}{L}
\label{Eq:Orowan}
\end{equation} 

\noindent where $\beta$ is a prefactor that depends on the dislocation line tension (and that is assumed equal to 1 if the line tension is independent of the dislocation character \citep{Nembach1997,Martin1998}), $G$ the shear modulus of the matrix, $b$  the modulus of the Burgers vector and $L$ the inter-precipitate distance  \citep{Orowan1948}.   This model was refined by Brown \citep{Brown1964} and Bacon \citep{Bacon1967} to take into account the effect of the dislocation self-stress, that reduces the CRSS required to bypass the precipitates due to the attraction between opposite dislocation segments bowing around the precipitate. They provided an expression for the CRSS (the so-called Bacon, Kocks and Scattergood (BKS) model) by fitting the results of DD simulations, which reads \citep{Bacon1973}:

\begin{equation}
\tau_c^{BKS}=A\frac{Gb}{L}\left[\ln\left(\frac{\bar{D}}{b}\right)+B \right]
\label{Eq:BKS}
\end{equation} 

\noindent where $A$ is a constant that depends on the dislocation character, $\bar{D}$ the harmonic mean of $L$ and the precipitate diameter $D$, $\bar{D}=(D^{-1}+L^{-1})^{-1}$, and $B=0.7$ is a constant to fit the model predictions to the results provided by DD simulations.  

These expressions were further extended to deal with random distributions of spherical precipitates \citep{Foreman1966,Kocks1966} and Bacon {\it et al.} \citep{Bacon1973} proposed the following expression for the CRSS including the effect of the dislocation arms interactions for random distributions of spherical precipitates  \citep{Nembach1997}:  

\begin{equation}
\tau_c^{BKS,rand}=\frac{1}{2\pi\sqrt{1-\nu}}\frac{Gb}{\langle L\rangle}  
\frac{\left[\ln\left(2\bar{D}/b\right)\right]^{3/2}}{\left[\ln\left(\langle L\rangle/b\right)\right]^{1/2}}
\label{Eq:BKS_random}
\end{equation}

\noindent where $\langle L\rangle$ stands for the average inter-precipitate distance and the constant  $ 1/(2\pi\sqrt{1-\nu})$ is the geometric mean of the constant $A$ for edge and screw dislocations. Further investigations extended these results to other precipitate geometries (disks, rods) which appear in different alloys \citep{Nie1996,Nie1998,Nie2003,Nie2008}. 
 
\begin{figure}
\centering
	\includegraphics[width=0.8\textwidth]{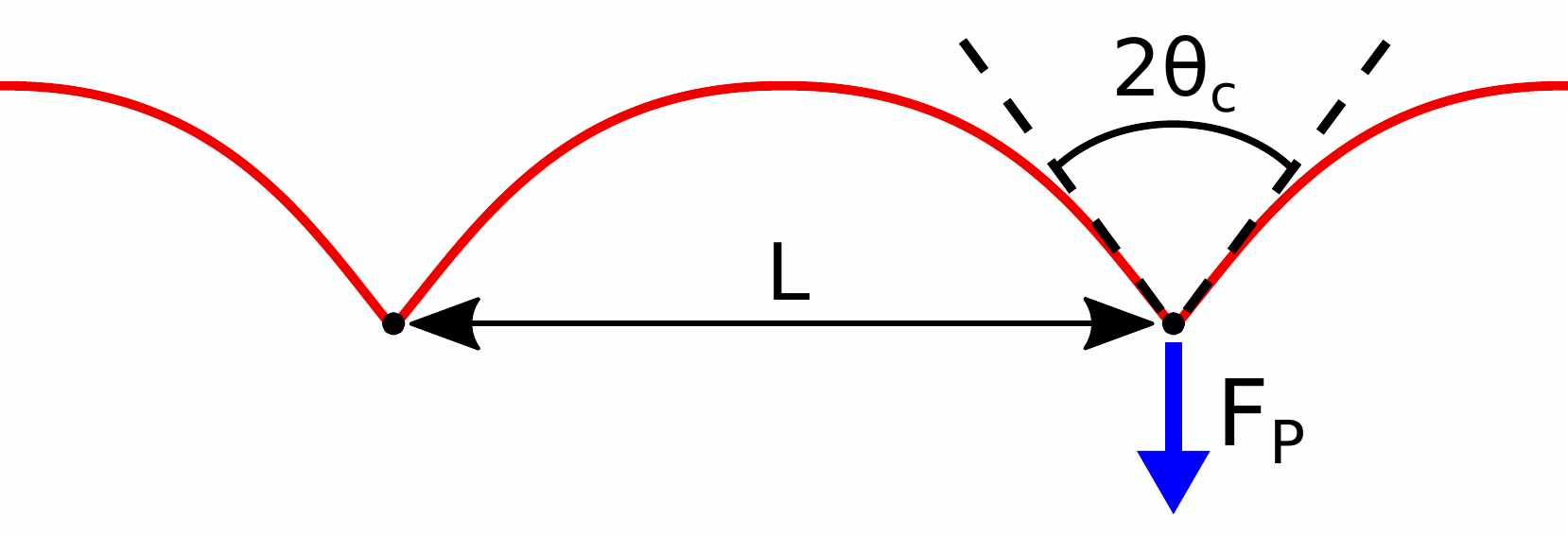}
	\caption{Schematic of a dislocation bowing around two point-like obstacles separated by a distance $L$. The dislocation exerts a force on the obstacles that increases as the angle $\theta$ between the dislocation arms decreases. The dislocation is sheared when this force reaches $F_P$, the precipitate strength.}
	\label{Fig:ObstacleAngle}
\end{figure}

Shearable precipitates were considered initially as point-like obstacles of finite strength. The precipitate strength was characterized by the maximum force that the obstacle can withstand before shearing, that is related to the angle formed by the dislocation arms around the precipitates (Fig. \ref{Fig:ObstacleAngle}). Friedel calculated the CRSS for a random distribution of point-like obstacles that can be sheared by dislocations as \citep{Friedel64}

\begin{equation}
\tau_c^{sh,rand}=\frac{Gb}{\langle L\rangle}\frac{\ln(\langle L\rangle/b)}{2\pi}\left(\cos(\theta_c)\right)^{3/2}
\label{Eq:OrowanShearRand}
\end{equation} 

\noindent where $\theta_c$ is the critical angle that defines the precipitate strength, $F_P$. This model was valid under the assumption of weak precipitates, $\theta_c >45^{\circ}$) \citep{Foreman1966}.

\begin{figure}
\centering
	\includegraphics[width=0.8\textwidth]{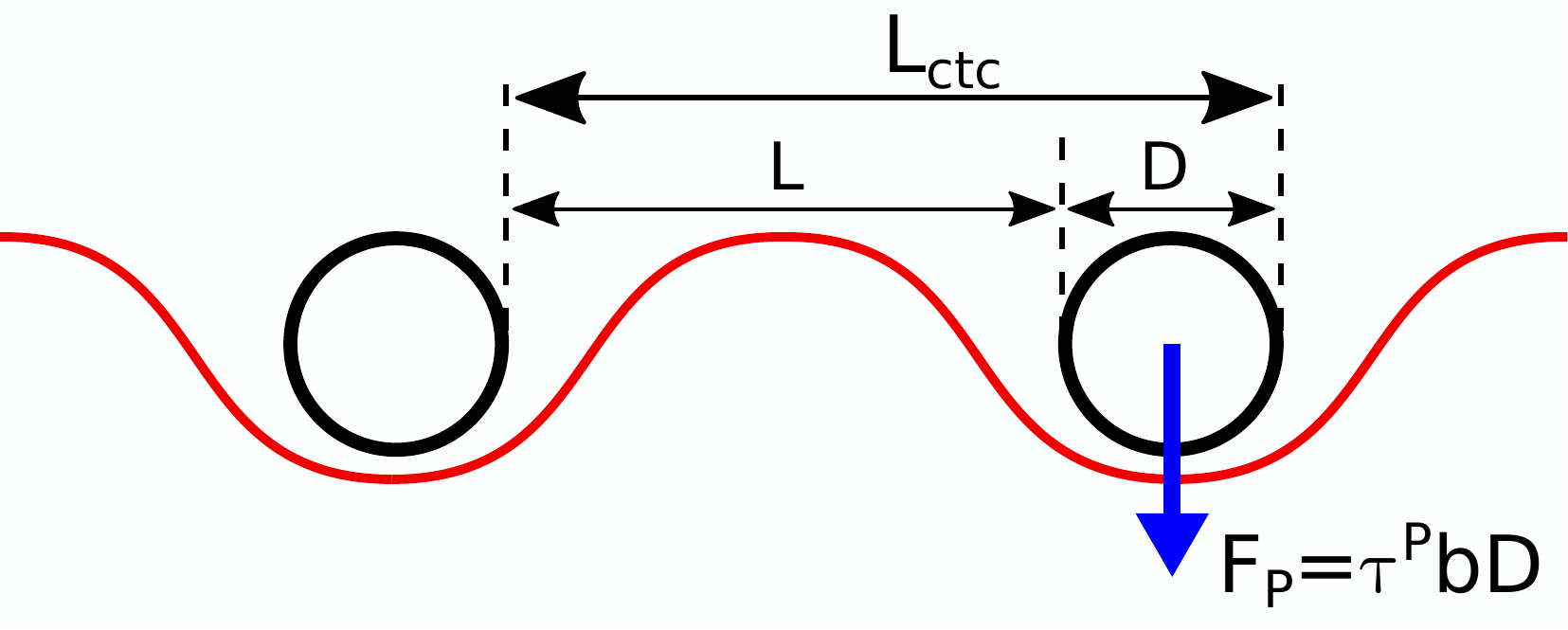}
	\caption{Schematic of a dislocation bowing around two finite size spherical obstacles of diameter $D$ in the slip plane separated by a distance $L$. The maximum force that the precipitate can withstand is related to the friction stress in the precipitate, $\tau^P$, according to $F_P = \tau^P b D $.}
	\label{Fig:ObstacleFriciton}
\end{figure}

Nevertheless, the approximation based on point-like obstacles cannot be extended to finite-size precipitates. Shearable finite-size precipitates were analyzed by Monnet \citep{Monnet2018} through the introduction of a friction stress to dislocation motion, $\tau^P$, that hinders the movement of the dislocation segments within the precipitate. The strength of the precipitate is then controlled by $\tau^P$, that accounts for the contributions of chemical, stacking fault and antiphase boundary energies (in the case of ordered precipitates)  as well as from  other sources of resistance to precipitate shearing (Peierls stress) \citep{Nembach1997}. Thus, the CRSS for a regular square distribution of spherical precipitates with diameter $D$  in the slip plane is obtained from the mechanical equilibrium between the dislocation driving force due to the applied shear stress and the resistance to precipitate shearing and it is given by (Fig. \ref{Fig:ObstacleFriciton}):

\begin{equation}
\tau_c^{sh}=\frac{\tau^P D}{L_{ctc}}
\label{Eq:ShearableAnalytical} 
\end{equation}

\noindent where $L_{ctc}$ is the centre-to-centre distance between precipitates along the dislocation line.  

These analytical expressions --based on line tension models-- do not take into account important factors that may change the CRSS, such as the elastic mismatch between the matrix and the precipitates. Moreover, the influence of the elastic mismatch may be different in the case of impenetrable or shearable precipitates \citep{Takahashi2008}. In addition, there are no analytical models that allow to determine accurately the effect of different parameters (precipitate size, volume fraction, strength) on the CRSS for both shearable and impenetrable precipitate distributions. This information will be very useful to design precipitate hardened alloys with optimum properties that are usually found in the regions where precipitate shearing and precipitate looping co-exist \citep{Bellon2020}.

Santos-G\"uemes {\it et al.} \citep{Santos-Guemes2020,Santos-Guemes2021} presented recently a multiscale DD framework that included the most important physical mechanisms that control dislocation/precipitate interactions, such as precipitate size, shape and spatial distribution, coherency strains and elastic mismatch. This strategy was successfully applied to two different  Al-Cu alloys containing either impenetrable \citep{Santos-Guemes2020} or shearable precipitates \citep{Santos-Guemes2021}, and the predictions of the CRSS were in excellent agreement with experimental results obtained from micropillar compression tests of single crystals oriented for single slip.  In this investigation, this simulation strategy is used to carry out a parametrical study of the effect of precipitate diameter and spacing on the CRSS for a regular array of spherical precipitates. Shearable and impenetrable precipitates are considered in the simulations and the effect of the elastic mismatch is accounted for with precipitates that are either stiffer or more compliant than the matrix. This information is used to develop a novel line tension model that is able to predict accurately the results of the DD simulations in all cases. Afterwards, the model is extended to deal with random distributions of spherical obstacles and its predictions are validated against new DD simulations. As a result, a novel generalized line tension model is presented that can take into account the effect of precipitate size, volume fraction and strength as well as of the elastic mismatch between matrix and precipitates on the CRSS for random precipitate distributions.

\section{Dislocation dynamics framework}

The dislocation dynamics framework in \cite{Santos-Guemes2021} was used in this investigation. Only the main features are presented below for the sake of completion and more details  can be found in \cite{Santos-Guemes2021} and \cite{Kohnert2021}. 

The dislocation lines are discretized into straight segments separated by nodes. The velocity of node $i$ in a glide plane is given by

\begin{equation} \label{Eq:mobility}
\mathbf{v}_i = \left\{
 \begin{array}{ll}
 [\mathbf{F}_i^{g}-F_i^{fric}(\mathbf{F}_i^{g}/|\mathbf{F}_i^{g}|)]/B & \mathrm{if} \quad |\mathbf{F}_i^{g}| > F_i^{fric}
\\ \\
 0 &  \mathrm{if} \quad|\mathbf{F}_i^{g}| \leq F_i^{fric}
 \end{array}
 \right.
\end{equation}
 
\noindent where $\mathbf{F}_i^{g}$ is the projection of the nodal force, $\mathbf{F}_i$, on the glide plane (characterized by the slip plane normal $\mathbf{n}$) according to

\begin{equation}
\mathbf{F}_{i}^{g} =  \mathbf{F}_{i} - (\mathbf{F}_{i} \cdot \mathbf{n})\mathbf{n}.
\end{equation}

The force on node $i$ was computed from the forces acting on its adjacent segments as

 \begin{equation}
\mathbf{F}_i = \sum_j \mathbf{f}_{ij}
\label{Fi}
\end{equation}

\noindent where $\mathbf{f}_{ij}$ is the force acting on the segment $ij$ (limited by nodes $i$ and $j$), which is computed according to

 \begin{equation}
\mathbf{f}_{ij} = \int_{\mathbf{x}_i}^{\mathbf{x}_j} N_i(\mathbf{x}) \mathbf{f}_{ij}^{pk}(\mathbf{x}) \rd \mathbf{x}
\label{fij}
\end{equation}

\noindent where $N_i$ is the interpolation function associated with node $i$ and $\mathbf{f}_{ij}^{pk}$ is the Peach-Koehler force given by

 \begin{equation}
\mathbf{f}_{ij}^{pk}(\mathbf{x}) =  \Big(\boldsymbol{\sigma}(\mathbf{x}) \cdot  \mathbf{b}_{ij}\Big) \times \mathbf{\hat t}_{ij} 
\label{fijpk}
\end{equation}

\noindent where $\boldsymbol{\sigma}(\mathbf{x})$ is the stress tensor at point $\mathbf{x}$, $\mathbf{b}_{ij}$ is the Burgers vector of the segment $ij$ and $\mathbf{\hat t}_{ij}$ the unit vector parallel to the dislocation line.

$F_i^{fric}$ in eq. \eqref{Eq:mobility} is a force threshold for dislocation glide. It is included in the nodes of the dislocation line within a shearable precipitate and represents the resistance to shearing of the precipitate. It  can be determined from eqs. \eqref{Fi}  to  \eqref{fijpk}  by assuming that the Peach-Koehler force in each dislocation segment that accounts for the precipitate resistance to shearing, $f^{pk,P}_{ij}$, is given by

\begin{equation}
f^{pk,P}_{ij}=\tau^{P}b
\end{equation}

\noindent where $\tau^P$ stands for the threshold shear stress necessary to shear the precipitates. It should be noted that the Burgers vector $b$  and the viscous drag coefficient $B$ in eq. \eqref{Eq:mobility} were assumed to be the same in the matrix and in the precipitate. The values for Al were used in these simulations, following \citep{Cho2017,Santos-Guemes2018}.

The stress field within the cubic simulation domain $V$, that determines dislocation slip through the Peach-Koehler force (eq. \eqref{fijpk}), was computed using an efficient FFT algorithm to solve the mechanical equilibrium equations with periodic boundary conditions
 
\begin{equation} \label{eq:MechEq}
 \left\{
 \begin{array}{l}
  \tensor{\sigma}(\mathbf{x})=\ftensor{C}(\mathbf{x}):[\tensor{\epsilon}(\mathbf{x})-\tensor{\epsilon}^p(\mathbf{x})], \quad \forall \mathbf{x} \in V
\\
 \mathrm{div} ( \tensor{\sigma}(\mathbf{x}))=0  \quad \mathbf{x} \in V \\
\tensor{\sigma}  \cdot \mathbf{n} \text{ opposite on opposite sides of } \partial V \\
\frac{1}{V}\int_V \tensor{\epsilon}(\mathbf{x})=E
 \end{array}
 \right.
\end{equation}\\

\noindent where $\ftensor{C}$ denotes the fourth order elasticity tensor, $\tensor{\epsilon}$ the total strain, $\tensor{\epsilon}^p$ the plastic strain and $\partial V$ stands for the boundaries of domain $V$ with normal $\mathbf{n}$ and $E$ is the imposed macroscopic strain. Note that the elasticity tensor is spatially dependent on the position $\mathbf{x}$, and therefore it may be different if $\mathbf{x}$ is located in the matrix or in a precipitate. The plastic strain $\tensor{\epsilon}^p$ was obtained from the Field Dislocation Mechanics method \citep{Djaka2017,Santos-Guemes2021,Kohnert2021}. Details of the FFT resolution can be found in \cite{Bertin2018}.

\section{Simulation details}

\begin{figure}[t!]
\centering
	\includegraphics[width=0.8\textwidth]{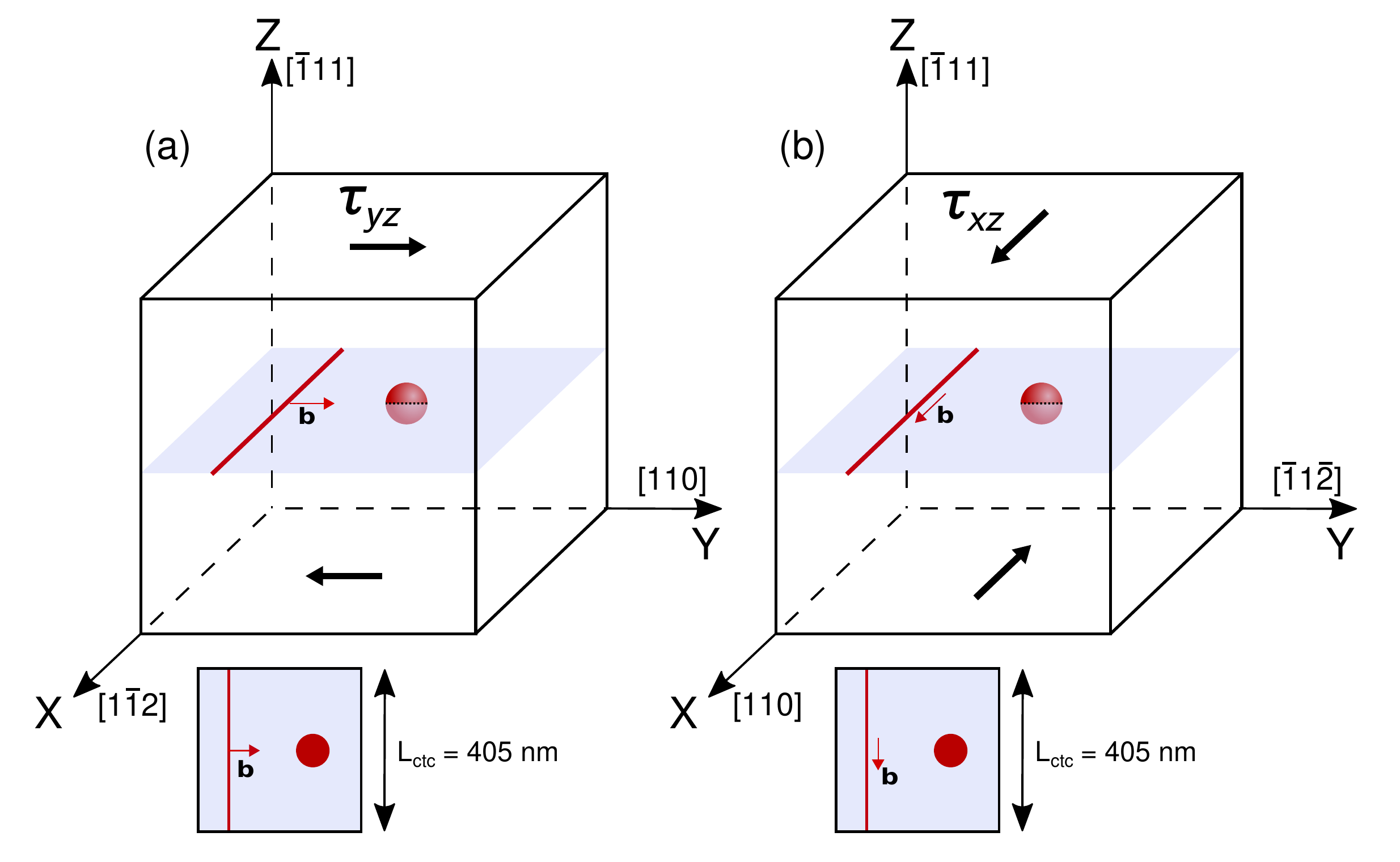}
	\caption{Initial configuration of the simulation domain including a spherical precipitate. (a) Initial edge dislocation and (b) initial screw dislocation. A slip plane section is shown below for both cases. The slip plane intersects the precipitate in the middle.}
	\label{Fig:InitialSphere}
\end{figure}

DD simulations were carried out using a cubic domain of 405 x 405 x 405 nm with periodic boundary conditions, which was discretized with a grid of 64 x 64 x 64 voxels. Both initial edge and screw dislocations were considered (Fig. \ref{Fig:InitialSphere}). The axes of the cubic domain were aligned with the [1$\overline{\mbox{1}}$2], [110] and [$\overline{\mbox{1}}$11] directions of the $\alpha$-Al fcc lattice for the case of a single straight edge dislocation ($\overline{\mbox{1}}$11)[110] (Fig. \ref{Fig:InitialSphere}a). In the case of a straight screw dislocation ($\overline{\mbox{1}}$11)[110], the axes of the cubic domain were aligned with the [110], [$\overline{\mbox{1}}$1$\overline{\mbox{2}}$] and [$\overline{\mbox{1}}$11] directions (Fig. \ref{Fig:InitialSphere}b). A shear strain rate of $5\cdot10^4 \rm{ s}^{-1}$ parallel to the ($\overline{\mbox{1}}$11) plane in the [110] direction was applied to the simulation domain (Fig. \ref{Fig:InitialSphere}). The initial configuration with a single straight dislocation in the box corresponds to a density of $6\cdot 10^{12}\rm{ m}^{-2}$. Previous studies have shown that the CRSS obtained from these DD simulations is independent of the strain rate for this dislocation/precipitate configuration when the strain rate is equal to or lower than  $5\cdot10^4 \rm{ s}^{-1}$ 
\citep{Santos-Guemes2018} and is equivalent to the one found under quasi-static conditions. The CRSS in the simulations corresponds to the peak stress in the shear stress - strain curve, that is, the stress that is required for the dislocation to glide through the entire simulation domain.

The properties of Al were chosen for the matrix. Thus, it was isotropic, with a shear modulus $G^M$ = 26.175  GPa, a Poisson ratio $\nu$ = 0.3 and Burgers vector $b$ =  0.286 nm oriented along the [110] direction.
 The shear modulus of the precipitate was varied proportionally to the shear modulus of the matrix according to $G^P=aG^M$. Thus, $a$ = 1 corresponds to homogeneous materials, whereas $a>1$ stands for stiffer precipitates and $a<1$ for more compliant precipitates. The size of the spherical precipitate in the simulation domain was also varied and,  hence, the distance between precipitates was changed accordingly. Different sets of simulations were performed considering  impenetrable precipitates and shearable precipitates, and the effect of the elastic heterogeneity was analyzed in both cases.

\section{Impenetrable precipitates}

\subsection{Homogeneous}

\begin{figure}
\centering
	\includegraphics[width=0.49\textwidth]{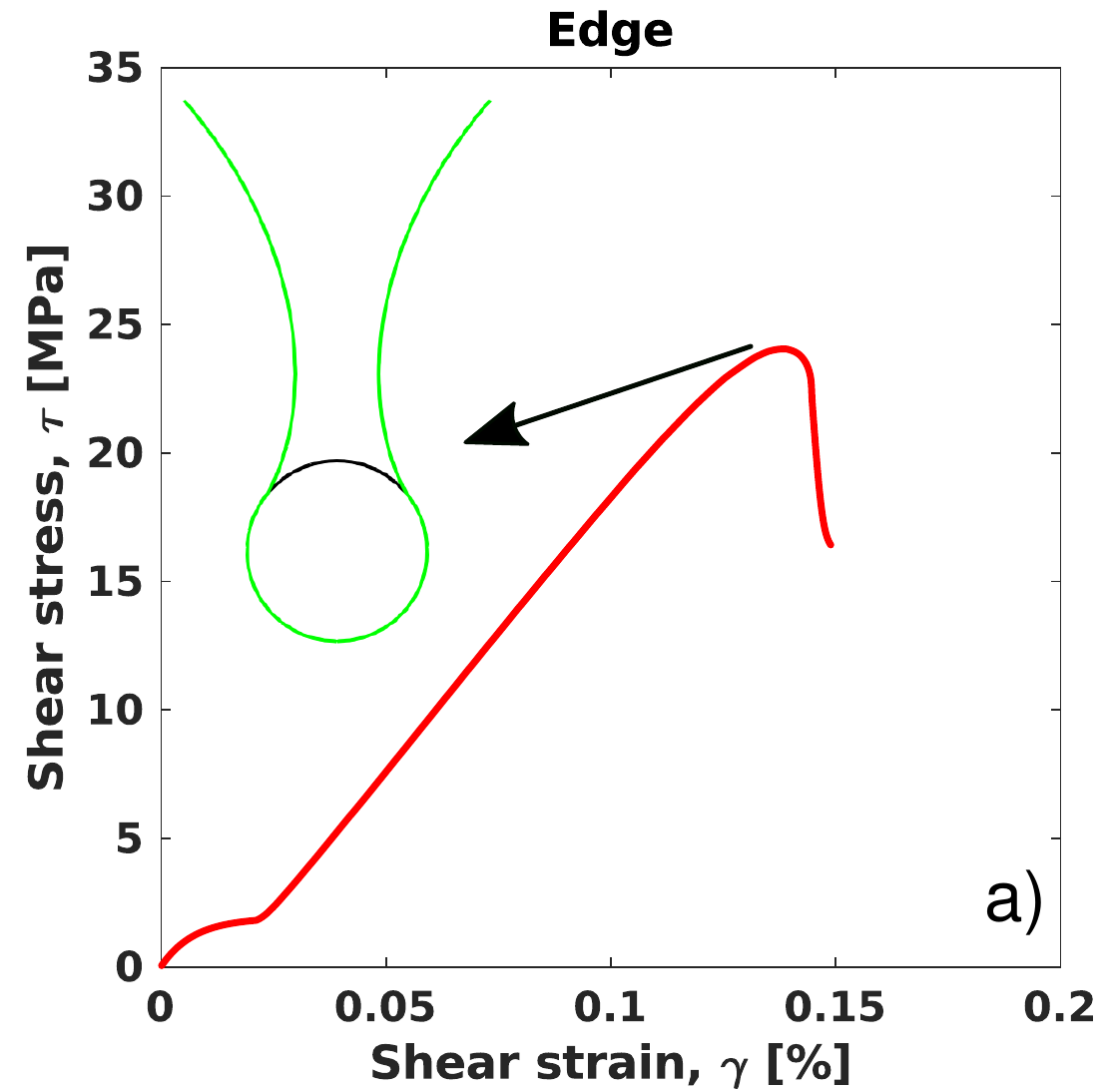}
	\includegraphics[width=0.49\textwidth]{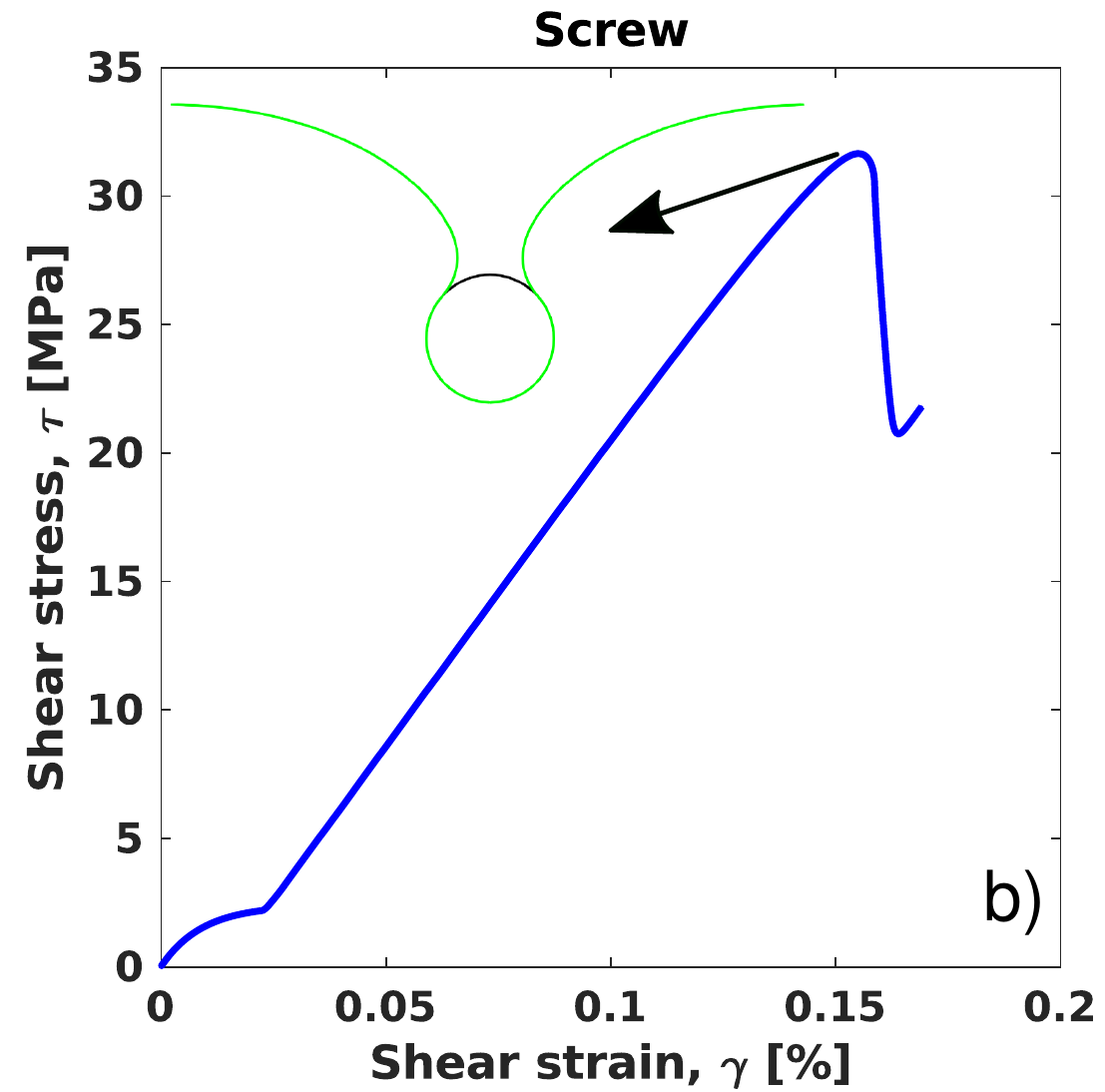}
	\caption{Shear stress - strain curve of the interaction of a dislocation with a precipitate of $D$ = 82 nm with the same elastic constants as the matrix.  (a) Edge dislocation. (b) Screw dislocation. A snapshot of the critical configuration at the CRSS is shown in each case.}
	\label{Fig:Curves_Imp_Homo}
\end{figure}

The interaction of a dislocation with a regular square distribution of impenetrable spherical precipitates with the same elastic constants as the matrix was studied first.  Spherical precipitates impenetrable to dislocations and with different diameters were introduced in the simulation domain, and they interacted with edge or screw dislocations. Two stress-strain curves obtained from the simulations with a precipitate of $D$ = 82 nm are depicted in Fig. \ref{Fig:Curves_Imp_Homo}, one for an edge dislocation (a) and another for a screw dislocation (b). A snapshot of the dislocation bowing around the precipitate in the critical configuration is also included in both cases. It can be observed that the edge dislocation  (Fig. \ref{Fig:Curves_Imp_Homo}a) tends to be aligned with the vertical direction, whereas the  screw dislocation (Fig. \ref{Fig:Curves_Imp_Homo}b) tends to be aligned with the horizontal axis, that coincides with the direction of the Burgers vector in each case.

\begin{figure}[!h]
\centering
	\includegraphics[width=0.49\textwidth]{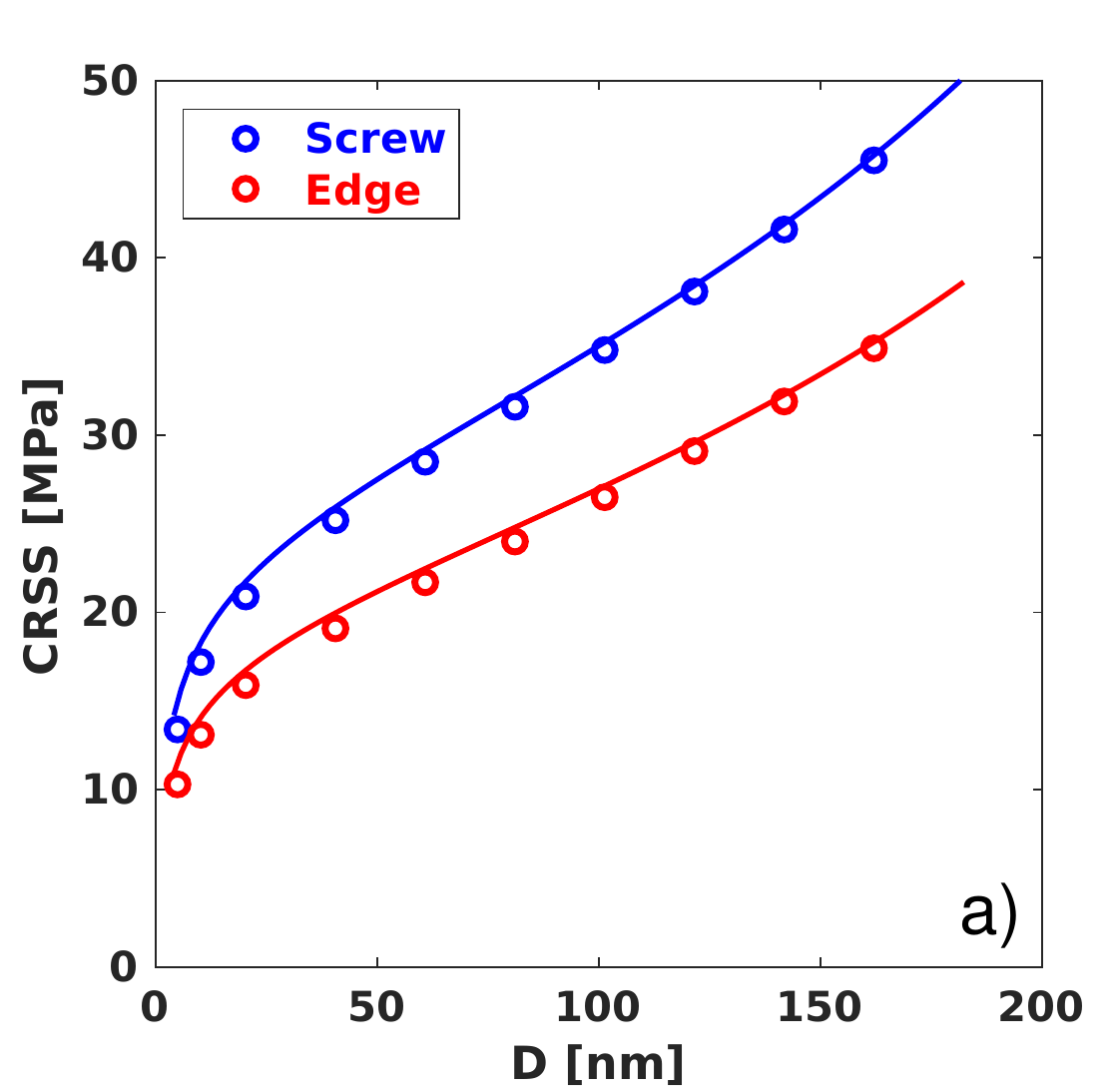}
	\includegraphics[width=0.49\textwidth]{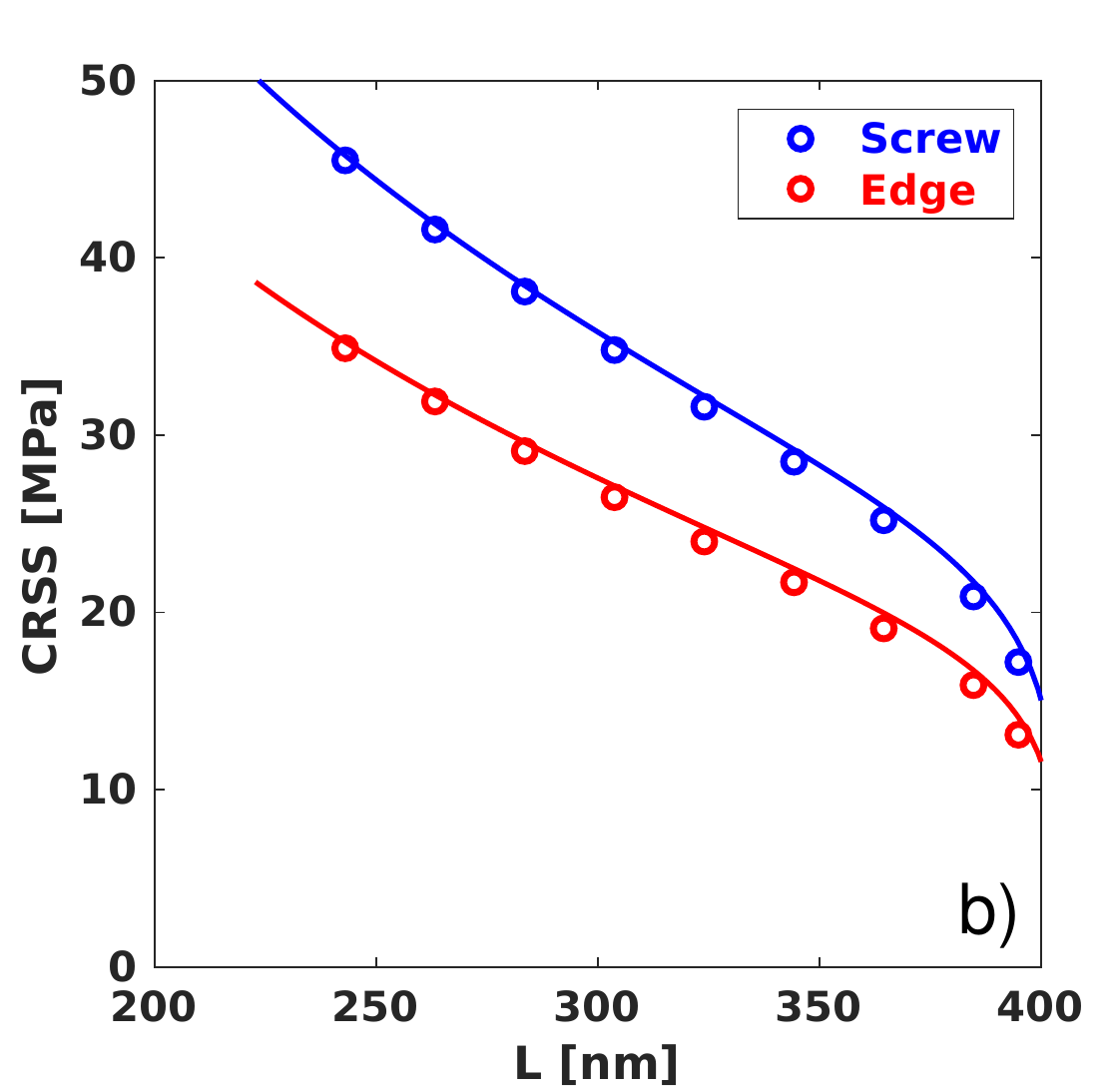}
	\caption{Predictions of the CRSS according to the BKS model (solid lines), eq. \eqref{Eq:BKS}, and to  DD simulations (circles) for edge and screw dislocations with a constant $L_{ctc}$ $(=L+D)$ of 405 nm. The CRSS is given as a function of (a) the precipitate diameter $D$ and (b) the inter-precipitate spacing $L$.}
	\label{Fig:BKS_DDD}
\end{figure}

This configuration has been widely studied using line tension models. In particular, the BKS model, eq. \eqref{Eq:BKS}, - that includes the effect of the self-stress of the dislocation -  is considered as the most accurate one \citep{Queyreau2010}.  The CRSSs obtained from the DD simulations are compared with the predictions of eq. \eqref{Eq:BKS} in Fig. \ref{Fig:BKS_DDD} as function of either the precipitate diameter (Fig. \ref{Fig:BKS_DDD}a) or the distance between precipitates (Fig. \ref{Fig:BKS_DDD}b). The prefactor $A$ in eq. \eqref{Eq:BKS} depends  on the dislocation line tension (and, hence, on the dislocation character)  and $A=1.1/(2\pi)$ for edge dislocations and $A=1/(2\pi(1-\nu))$ for screw dislocations because line tension of screw dislocations is larger than that of edge dislocations. $\nu$ stands for  the Poisson's ratio of the metallic matrix. Thus, the CRSS for screw dislocations are larger than that for edge dislocations for a given precipitate size. Furthermore, the difference of the line tension favours the alignment of dislocation segments parallel to the Burgers vector (screw character) rather than perpendicular to it (edge character), and this influences the shape of the bowing out of the dislocation around the precipitate. The results in Fig.  \ref{Fig:BKS_DDD} show very good agreement between the BKS model and the DD simulations. Thus,  the BKS model is able to capture the abrupt reduction of the CRSS for small precipitates due to the attraction between opposite dislocation arms. On the contrary, the attraction of opposite dislocation arms plays a minor role on the CRSS in the case of large precipitates.

\subsection{Heterogeneous}

DD simulations were performed with different values of the shear modulus of the precipitate, $G^P$, while the Poisson's ratio was kept constant. Simulations were performed in the range $\frac{1}{3}G^M < G^P < 3G^M$, which covers the elastic mismatch between matrices and precipitates in most alloys of technological interest. The presence of an elastic heterogeneity in the simulation domain leads to a modification in the stress field, particularly in the region close to the precipitate, that affects the movement of the dislocation line. This modification is the result of two interactions: the stress field of the dislocation with the precipitate and the external load with the heterogeneous phase. According to the literature, the influence of the former (that results in the image stresses) is larger than the latter one \cite{Szajewski2021}. Both contributions are directly taken into account in the computation of the mechanical fields described in the previous section.

\begin{figure}
\centering
	\includegraphics[width=0.49\textwidth]{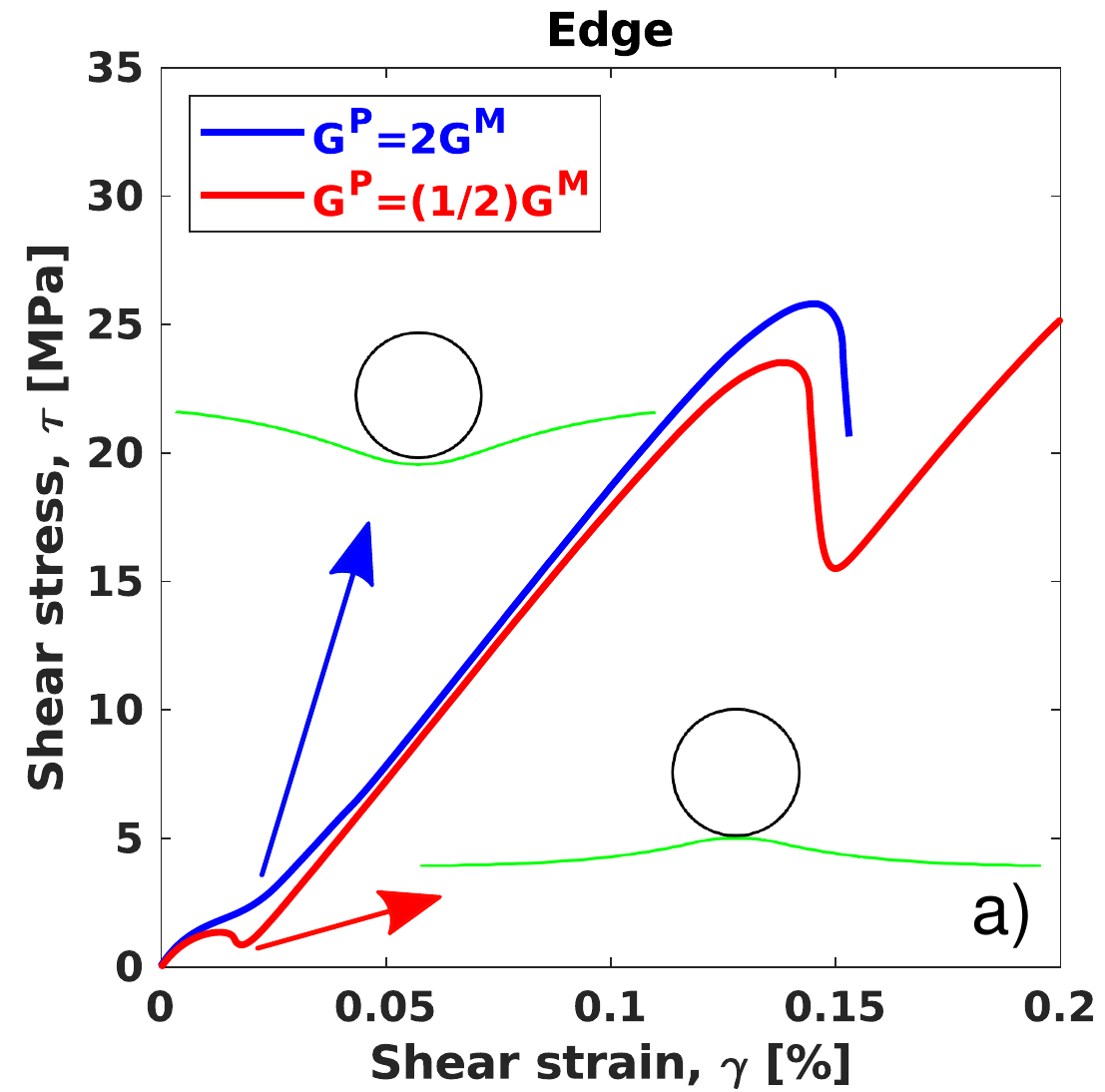}
	\includegraphics[width=0.49\textwidth]{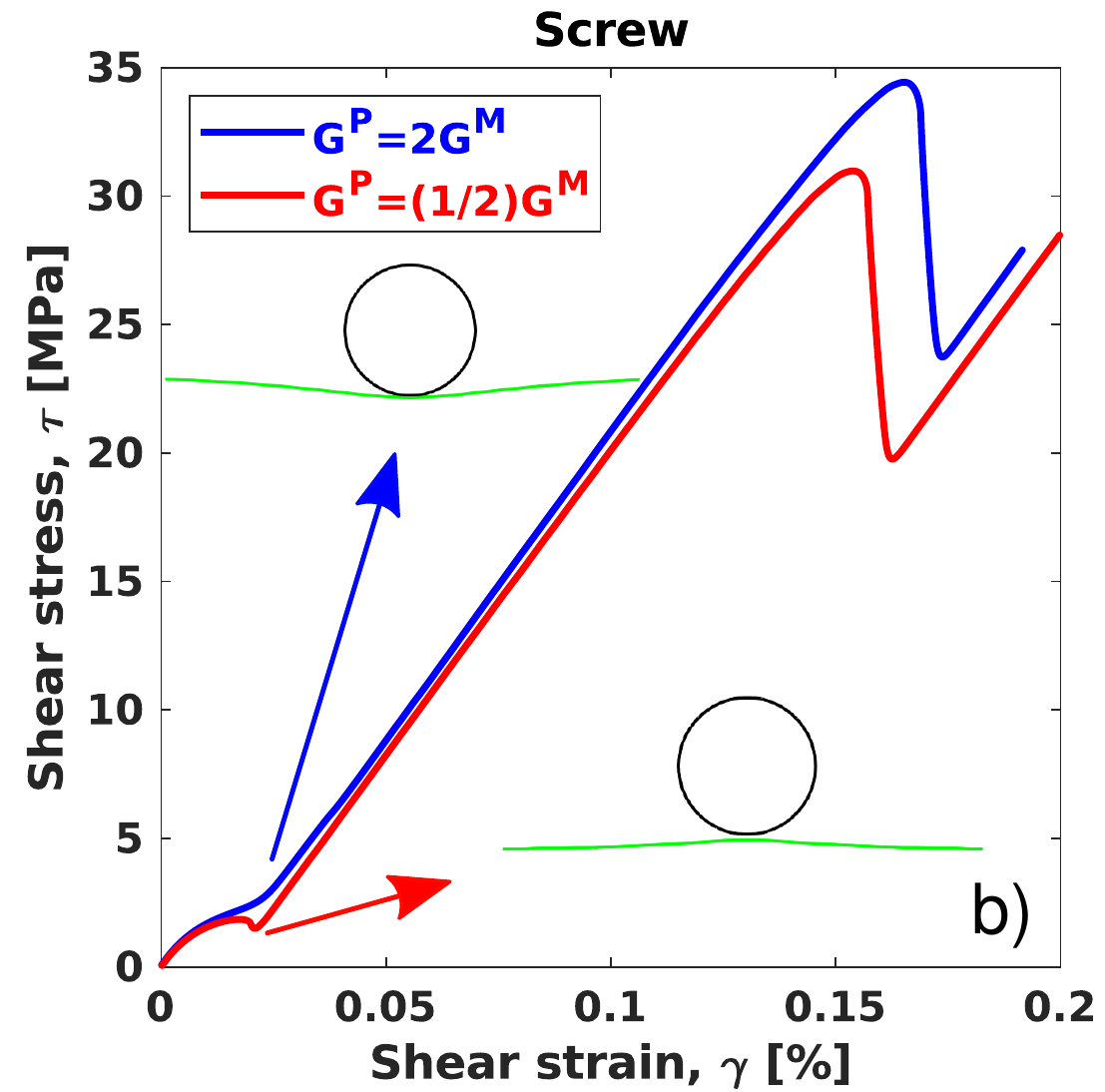}
	\caption{Shear stress - strain curves of the interaction of a dislocation with a precipitate with $D$ = 82 nm. The elastic constants of the precipitate are either twice the elastic constants of the matrix (blue lines) or half (red lines). A snapshot of the configuration when the dislocation is approaching the precipitate, indicating the effect of the image stresses, is shown in each case. (a) Edge dislocation. (b) Screw dislocation.}
	\label{Fig:Curves_Imp_Hetero}
\end{figure}

The stress-strain curves from simulations with a precipitate of $D$ = 82 nm stiffer than the matrix ($G^P=2G^M$, blue lines) and a precipitate more compliant than the matrix ($G^P=(1/2)G^M$, red lines) are plotted in Fig. \ref{Fig:Curves_Imp_Hetero}. The effect of the image stresses can be clearly observed when the dislocation approaches the precipitate (see snapshots included in Fig. \ref{Fig:Curves_Imp_Hetero}). Stiffer precipitates result in a repulsion of the dislocation line, increasing the shear stress required to move the dislocation towards the precipitate. On the contrary, softer precipitates attract the dislocation line, leading to a small reduction in the shear stress until the dislocation touches the precipitate. The attraction/repulsion of the dislocation continues during the bowing out of the dislocation around the precipitate, resulting in larger values of the CRSS for stiff precipitates than for soft precipitates for a given precipitate diameter.

\begin{figure}
\centering
	\includegraphics[width=0.49\textwidth]{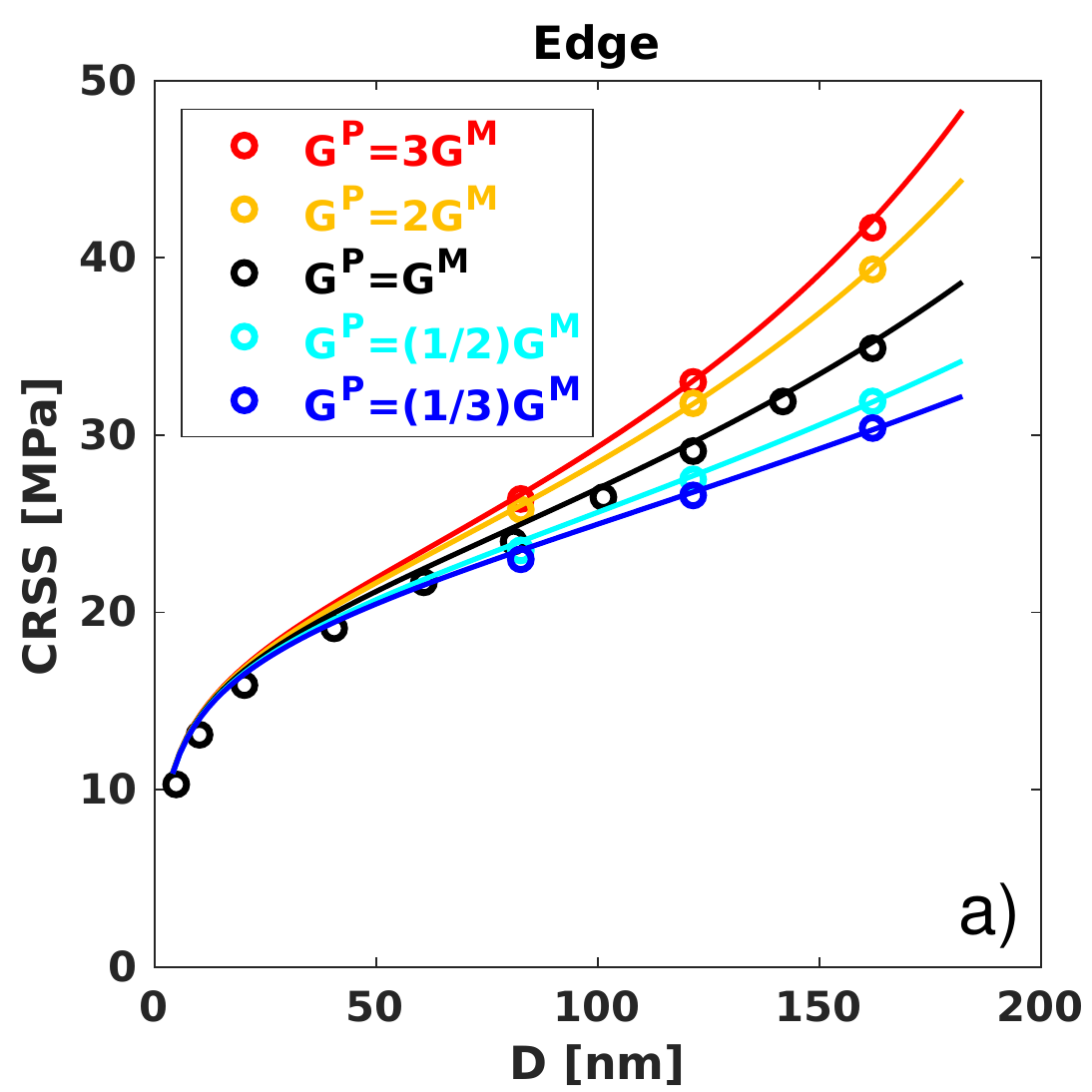}
	\includegraphics[width=0.49\textwidth]{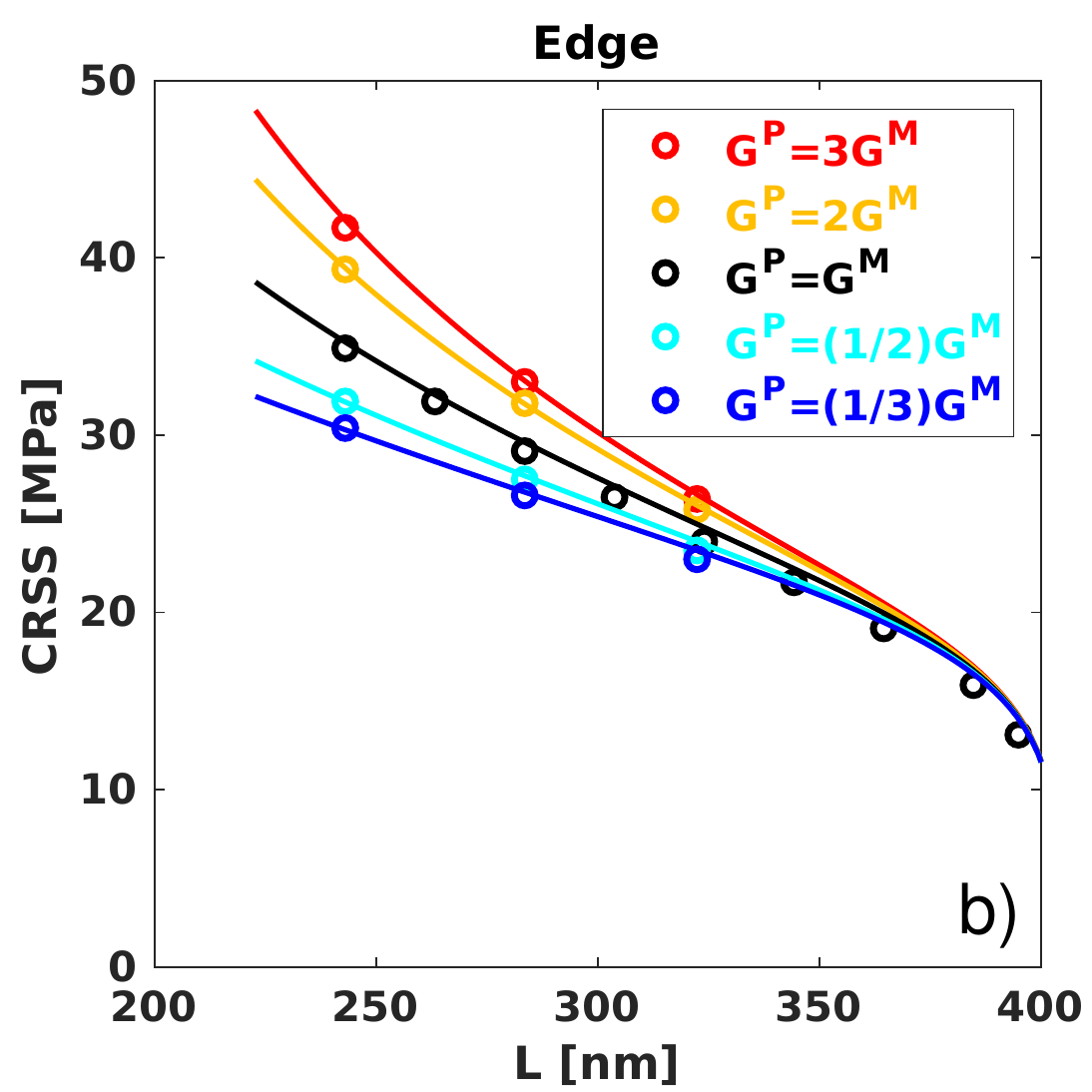}
	\caption{Comparison of the predictions of the CRSS for impenetrable and heterogeneous precipitates according to  eq. \eqref{Eq:BKS_Heterogeneous}  (solid lines), with the results of DD simulations for edge dislocations with a constant $L_{ctc}$ $(=L+D)$ of 405 nm. The different colours represent the different elastic constants of the precipitate with respect to the matrix ($G^P = a G^M$). The CRSS is given as a function of (a) the precipitate diameter $D$ and (b) the inter-precipitate spacing $L$. }
	\label{Fig:BKS_Edge}
\end{figure}

\begin{figure}
\centering
	\includegraphics[width=0.49\textwidth]{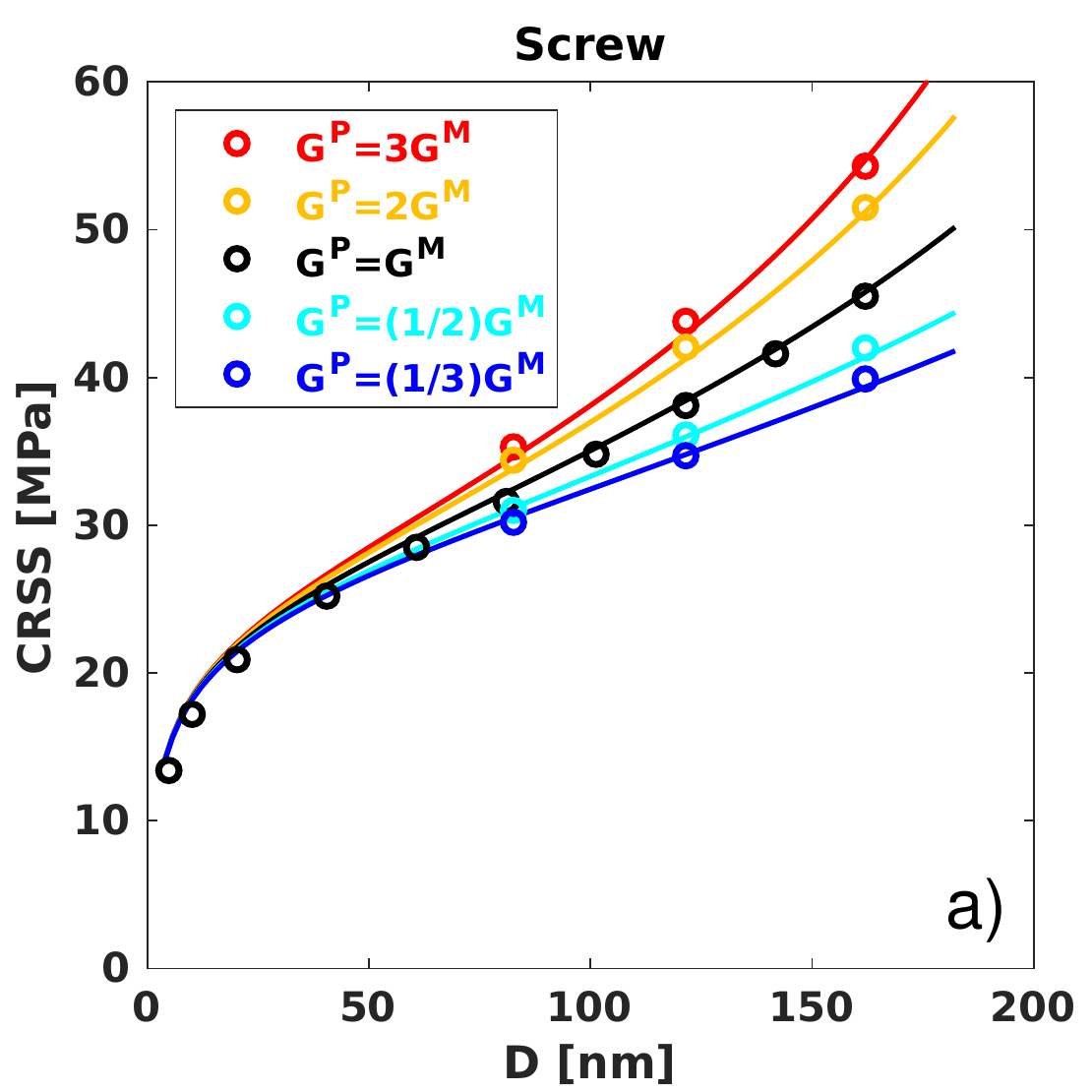}
	\includegraphics[width=0.49\textwidth]{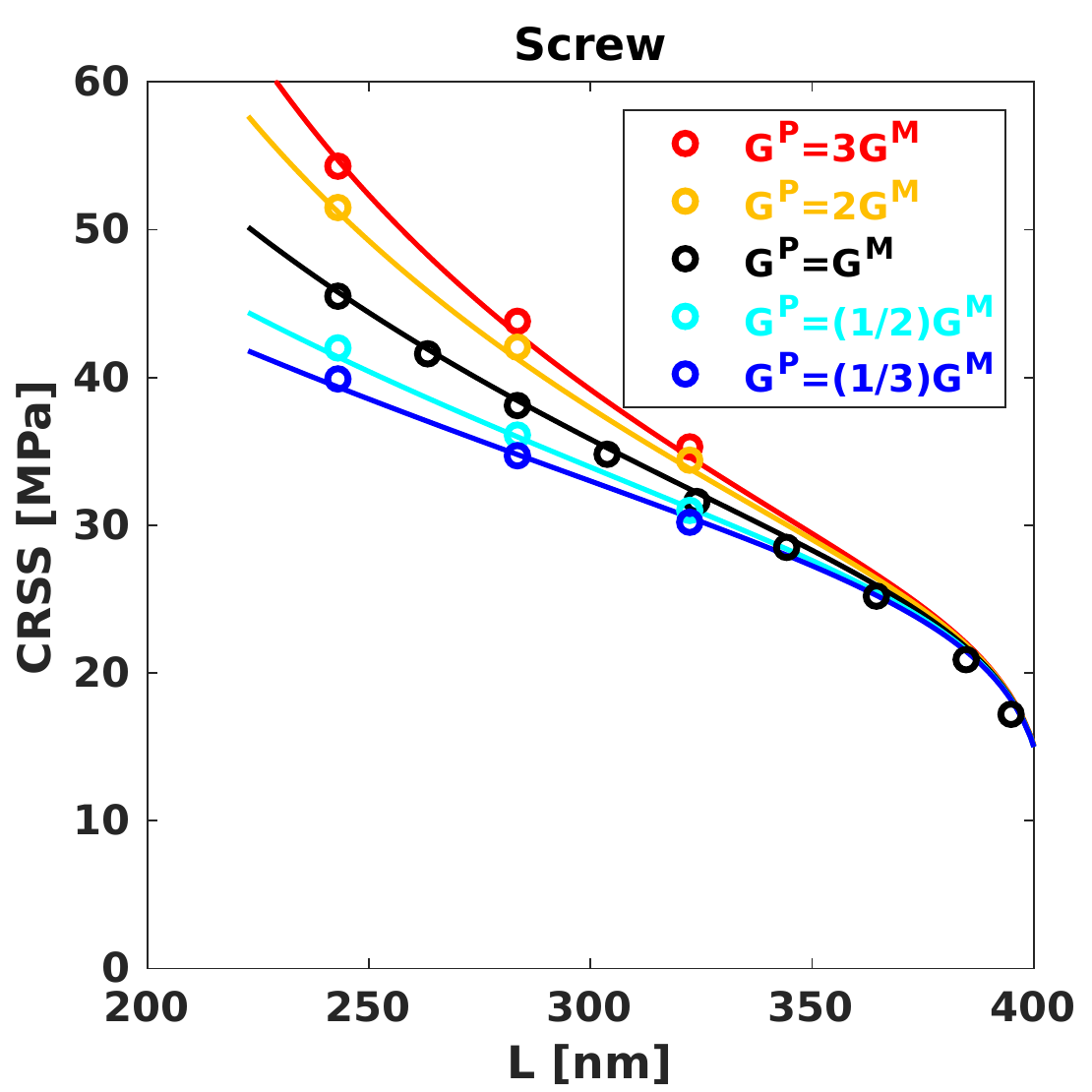}
	\caption{Comparison of the predictions of the CRSS for impenetrable and heterogeneous precipitates according to  eq. \eqref{Eq:BKS_Heterogeneous}  (solid lines), with the results of DD simulations for screw dislocations with a constant $L_{ctc} (=L+D)$ of 405 nm. The different colours represent the different elastic constants of the precipitate with respect to the matrix ($G^P = a G^M$). The CRSS is given as a function of (a) the precipitate diameter $D$ and (b) the inter-precipitate spacing $L$. }
	\label{Fig:BKS_Screw}
\end{figure}

It was found that the effect of elastic heterogeneity can be accounted for in the BKS model assuming that the effect of the image stresses is equivalent to a modification of the interprecipitate distance $L$. Therefore, an effective distance, $L_{eff}$, is introduced in eq. \eqref{Eq:BKS}, leading to 

\begin{equation}
\tau_c^{imp}=\frac{G^Mb}{L_{eff}}A\left[\ln\left(\frac{\bar{D}}{b}\right)+B \right]
\label{Eq:BKS_Heterogeneous}
\end{equation}

\noindent where 

\begin{equation}
L_{eff}=L\left(1-\frac{D}{L}\alpha\right)
\label{Eq:Leff}
\end{equation}

\noindent and  $\alpha$ is a factor that depends on the shear moduli of the matrix and the precipitate, $G^M$ and $G^P$, respectively, according to

\begin{equation}
\alpha=\rm{sign}(\Delta G)\frac{9}{20}\left(1-\min\left(\frac{G^P}{G^M},\frac{G^M}{G^P}\right)\right)^{3/2}
\label{Eq:AlphaBKS}
\end{equation}

\noindent where $\Delta G = G^P - G^M$. This phenomenological expression is valid for edge and screw dislocations and was selected to represent accurately the DD results in a wide range of elastic mismatches, keeping a relatively simple form. Obviously,  the form of the original BKS model, eq. \eqref{Eq:BKS} is recovered when $G^M = G^P$. The predictions of the CRSS from the DD simulations and from the modified BKS model, eq. \eqref{Eq:BKS_Heterogeneous},  are shown in Figs. \ref{Fig:BKS_Edge} and \ref{Fig:BKS_Screw} for edge and screw dislocations, respectively, for different values of $G^P$ and they are in close agreement.

The image stresses arising due to heterogeneity pushed the dislocation away from the precipitate in the case of stiffer precipitates, reducing the effective inter-precipitate distance. On the contrary, the image stresses attracted the dislocation towards the precipitate when it was more compliant, increasing the value of $L_{eff}$. These mechanisms lead to an increase in the CRSS for stiffer precipitates and to a reduction in the CRSS for more compliant precipitates.  The modified version of the BKS model is able to reproduce the results of the DD simulations for initial edge and screw dislocations interacting with a heterogeneous precipitate, both in the case of stiffer and more compliant precipitates. It can be observed that the effect of the elastic heterogeneity increases with the elastic mismatch and the size of the precipitate, being negligible for small precipitates.

\section{Shearable precipitates}

\subsection{Homogeneous}

The CRSS to cut a shearable and homogeneous precipitate can be determined from the mechanical equilibrium between the driving force associated with the applied shear stress on the glide plane and the resistance of the precipitate to shearing according to eq. \eqref{Eq:ShearableAnalytical} \citep{Monnet2018}. This expression only includes information about  geometrical parameters and the friction stress in the precipitate but it does not incorporate information about the elastic properties or the dislocation character. Therefore, the CRSS predicted is the same for edge and screw dislocations.

\begin{figure}
\centering
	\includegraphics[width=0.49\textwidth]{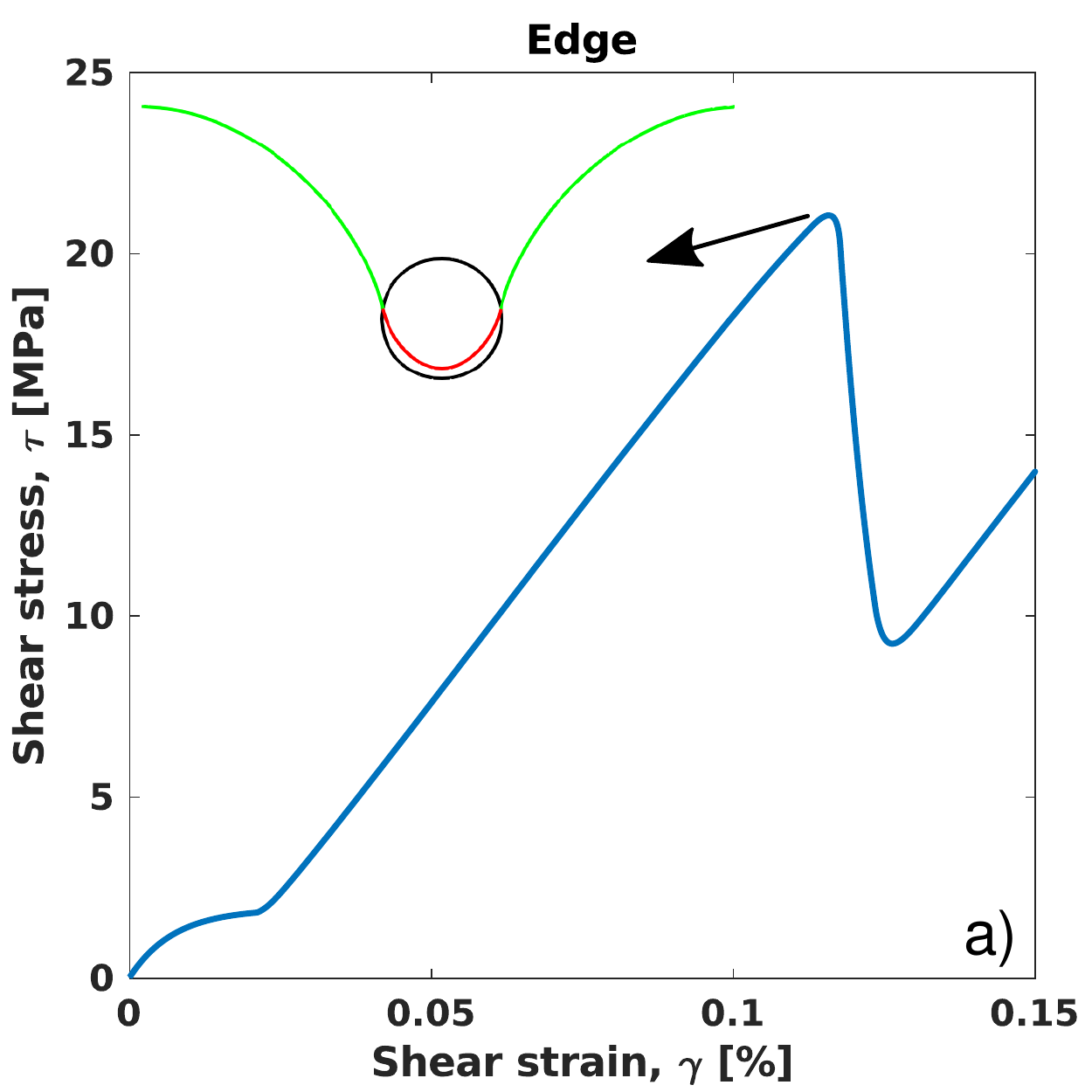}
	\includegraphics[width=0.49\textwidth]{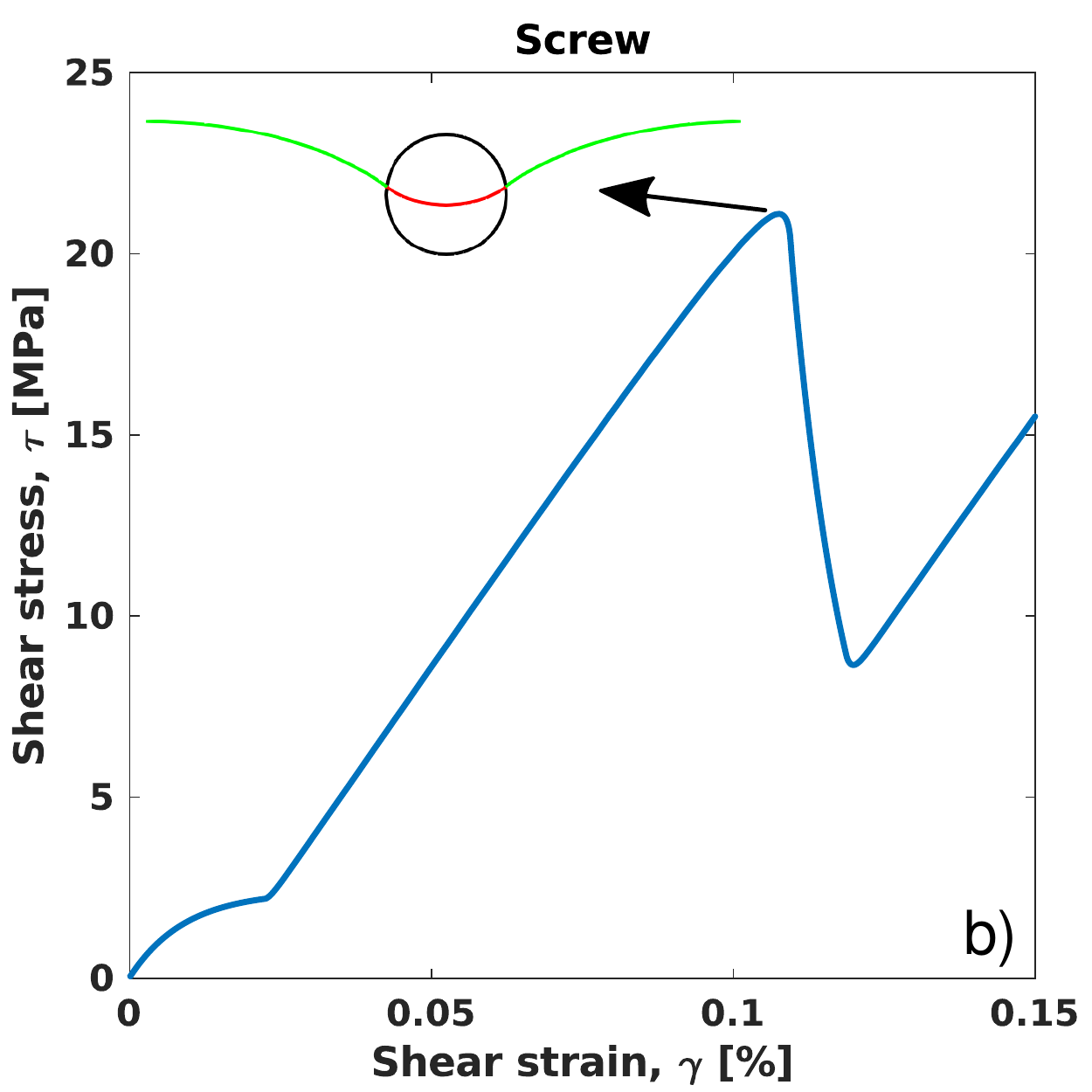}
	\caption{Shear stress - strain curves of the interaction of a dislocation with a shearable precipitate of $D$ = 82 nm with the same elastic constants as the matrix. The friction stress of the precipitate is $\tau^P$ = 100 MPa. A snapshot of the critical configuration at the CRSS is shown in each case. Green dislocation segments are located in the matrix while red segments are within the precipitate.  (a) Edge dislocation. (b) Screw dislocation.}
	\label{Fig:Curves_Shearable_Fric_Homo}
\end{figure}

The shear stress-strain curves obtained from DD simulations of the interaction of an edge and a screw dislocation with a shearable precipitate of $D$ = 82 nm and $\tau^P$ = 100 MPa are plotted in Figs. \ref{Fig:Curves_Shearable_Fric_Homo}a and b, respectively.  The snapshots of the critical configuration are also plotted for each case. The CRSS  is the same in both cases, even though the shapes of the dislocation line shearing the precipitate are different due to the differences in line tension. This result confirms that the dislocation character does not influence the CRSS in the case of shearable and homogeneous precipitates.

\begin{figure}
\centering
	\includegraphics[width=0.49\textwidth]{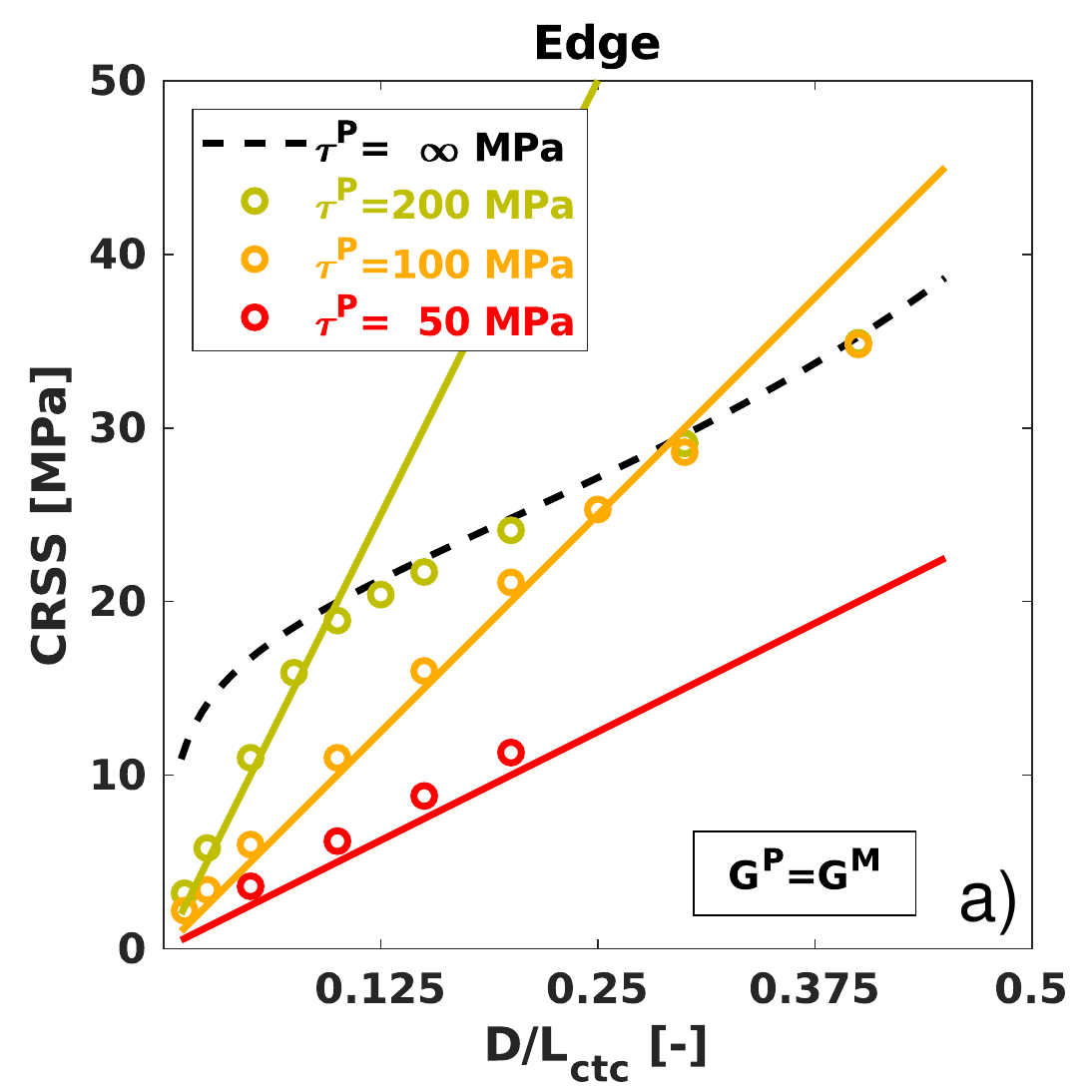}
	\includegraphics[width=0.49\textwidth]{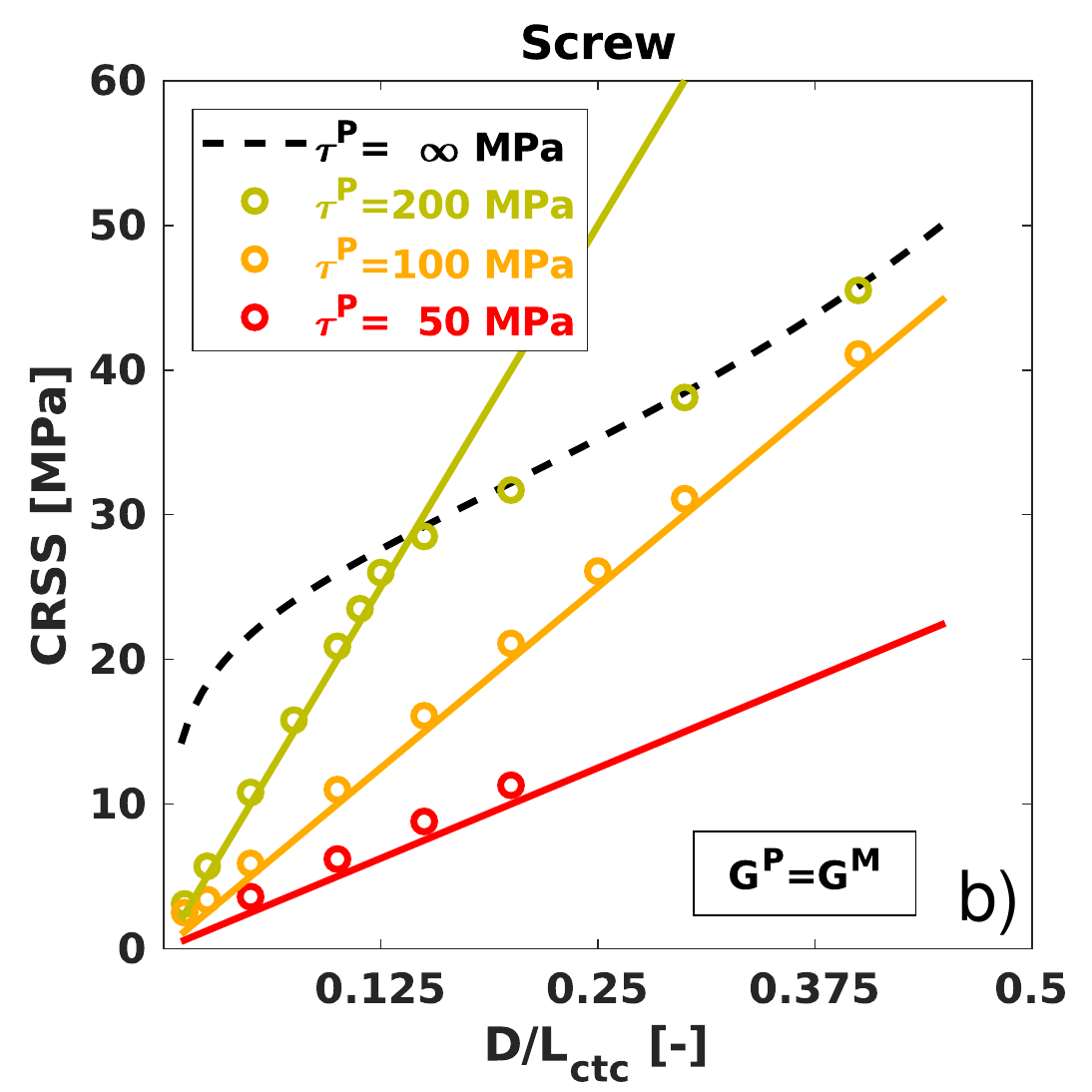}
	\caption{Comparison of the predictions of the CRSS for shearable and homogeneous precipitates according to  eq. \eqref{Eq:ShearableAnalytical}  (solid lines) with the results of DD simulations  for different values of $\tau^P$. The dashed black line represents the case of impenetrable precipitates.  (a) Edge dislocation. (b) Screw dislocation.}
	\label{Fig:Curves_Shearable}
\end{figure}

DD simulations were performed to assess the validity of eq. \eqref{Eq:ShearableAnalytical} to estimate the effect of the precipitate diameter and friction stress in the CRSS. The CRSSs predicted by the simulations are plotted in Fig. \ref{Fig:Curves_Shearable}a and b as a function of $D/L_{ctc}$ for edge and screw dislocations, respectively. The predictions of the CRSS from eq. \eqref{Eq:ShearableAnalytical} for sherable precipitates and from eq. \eqref{Eq:BKS} for impenetrable precipitates are also plotted in these figures. The agreement in the prediction of the CRSS between DD and line tension models is very good and confirms that the CRSS of shearable precipitates is proportional to the  friction stress and to the precipitate diameter. Furthermore,  the DD simulations show the transition between the shearable and impenetrable regimes that is captured by the intersection between the analytical models for impenetrable and shearable precipitates. Thus, shearable precipitates behave as impenetrable ones if the friction stress and/or the diameter are high enough. The dislocation is not able to shear completely the precipitate that is finally overcome by the formation of an Orowan loop.

It should be noted that the CRSS in these simulations refers to the maximum shear stress required to overcome the precipitate by the first dislocation. If the first dislocation shears the precipitate, subsequent dislocations (that appear in the simulations due to the periodic boundary conditions) also shear the precipitate leading to the same value of the CRSS. However, an Orowan loop is formed around the precipitate if the first dislocation cannot shear the precipitate. The next dislocation that approaches the precipitate is repelled by the Orowan loop and, thus, the CRSS to overcome the precipitate increases, leading to strain hardening.  This process is repeated with subsequent dislocations, leading to the formation of a  dislocation pile-up until the driving force in the inner loop is large enough to overcome $\tau^P$ and the precipitate is sheared. This behavior is in agreement with experimental evidence in $\theta'$ precipitates in Al-Cu alloys \citep{Kaira2018} or $\beta'_1$ precipitates in Mg-Zn alloys \citep{Alizadeh2020}. In both cases,  the precipitates are initially impenetrable to dislocations but they are finally sheared due to the formation of dislocation pile-ups after enough deformation is applied.

\subsection{Heterogeneous}

\begin{figure}[t]
\centering
	\includegraphics[width=0.49\textwidth]{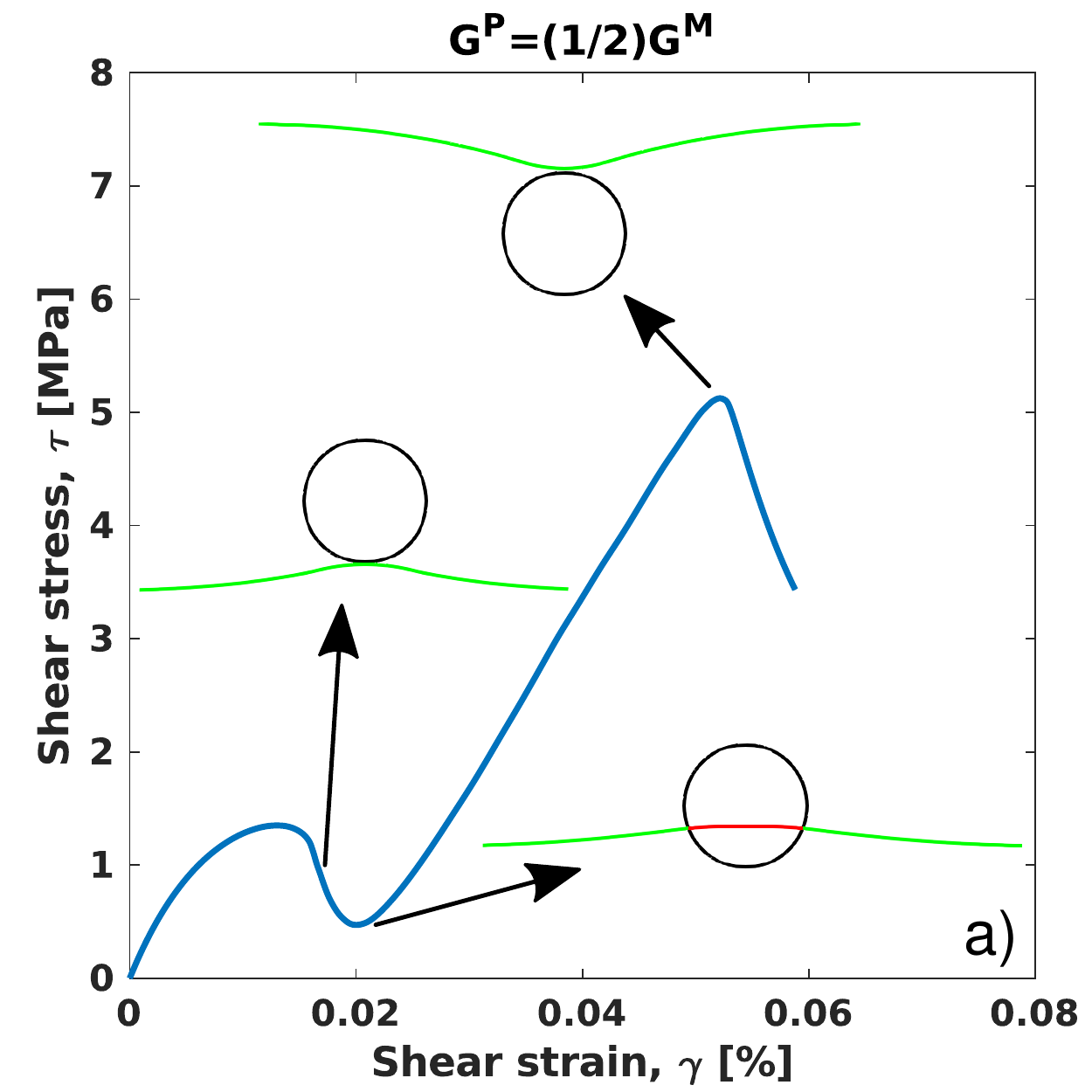}
	\includegraphics[width=0.49\textwidth]{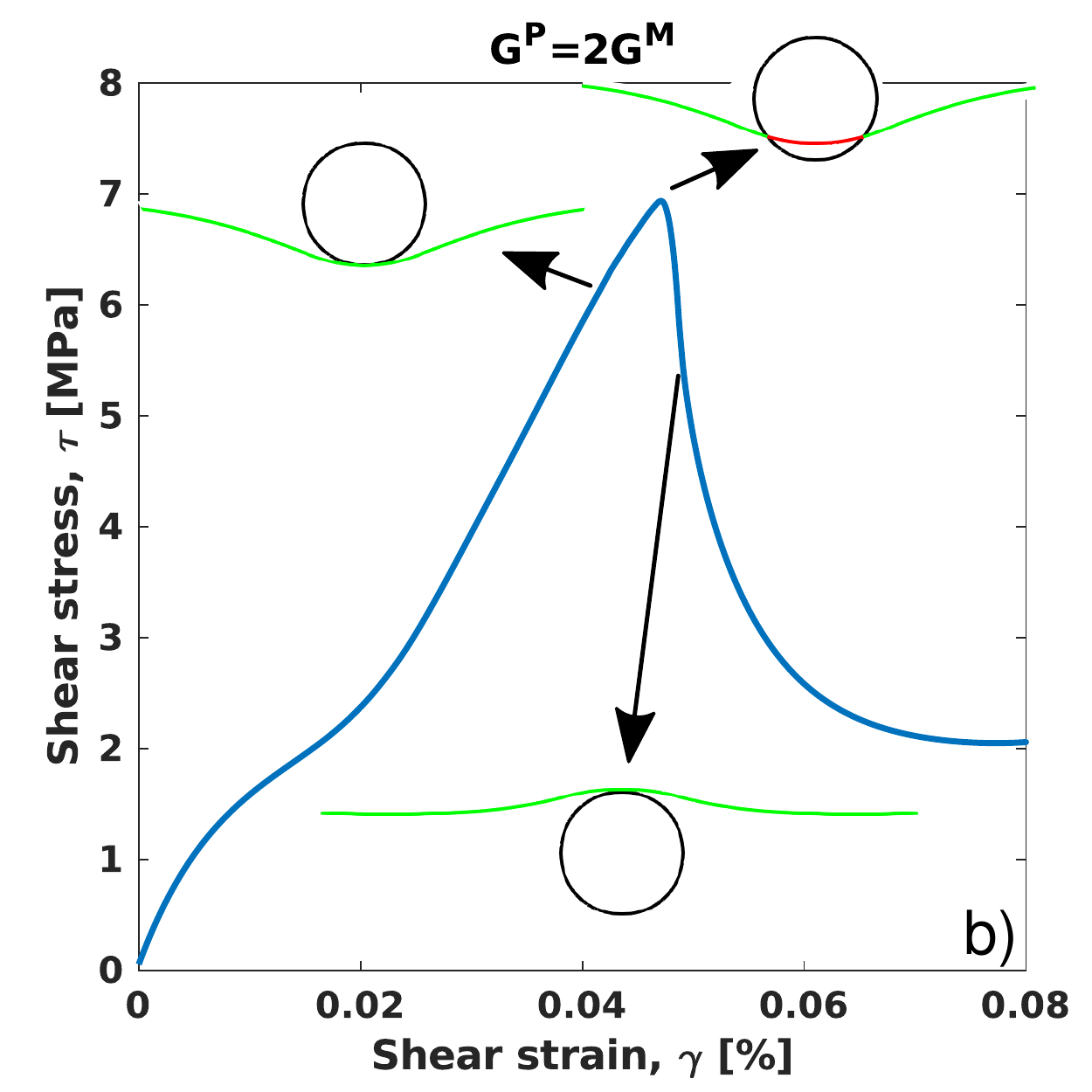}
	\caption{Shear stress - strain curves of the interaction of an edge dislocation with a shearable precipitate of $D$ = 82 nm whose elastic constants are different from those of the matrix. The  friction stress to shear the precipitate is $\tau^P$ = 0. Snapshots of the evolution of the dislocation are shown in each case. Green dislocation segments are located in the matrix and red segments are inside the precipitate.  (a) $G^P = 0.5 G^M$ (b) $G^P = 2 G^M$. }
	\label{Fig:Curves_Shearable_NoFric_Hetero}
\end{figure}

The influence of the elastic mismatch between the precipitate and the matrix on the CRSS for shearable precipitates is more complex  than for impenetrable ones because it depends on whether the dislocation segments are inside or outside the precipitate. When the dislocation line is outside the precipitate, the dislocation is repelled by precipitates that are stiffer than the matrix  and attracted by those that are more compliant than the matrix. When the dislocation line is inside the precipitate, the driving force for dislocation slip increases (with respect to that in the matrix) if the precipitate is stiffer than the matrix and decreases in the opposite case. The influence of these mechanisms on the CRSS can be firstly assessed by analyzing the interaction of a dislocation with a shearable precipitate whose friction stress is $\tau^P = 0$. In this case, the propagation of the dislocation within the precipitate is only controlled by the image stresses due to the heterogeneity.  The shear stress-strain curves resulting form the interaction of an edge dislocation with a shearable precipitate of $D$ = 82 nm and $\tau^P$ = 0 are plotted in Figs. \ref{Fig:Curves_Shearable_NoFric_Hetero}a and b when the precipitate is stiffer  or more compliant than the matrix, respectively. The initial shape of the shear stress-strain curve is identical to the one found for impenetrable precipitates: the dislocation line is attracted by the more compliant precipitate and repelled by the stiffer one. However, the dislocation/precipitate interactions in both cases are different when the dislocation begins to shear the precipitate. In the case of a softer precipitate (Fig. \ref{Fig:Curves_Shearable_NoFric_Hetero}a), the dislocation penetrates the precipitate until a minimum is reached in the stress-strain curve because the driving force on the dislocation segments within the precipitate (red segments in Fig. \ref{Fig:Curves_Shearable_NoFric_Hetero}a) is reduced due to the lower shear modulus of the precipitate. Thus,   the applied shear stress has to be increased to  shear completely the precipitate. On the contrary, most of the hardening provided by the stiffer precipitate is attained before the dislocation penetrates the precipitate because of the repulsion induced by the image stresses (Fig. \ref{Fig:Curves_Shearable_NoFric_Hetero}b). When the dislocation starts shearing the precipitate, the driving force on the dislocation segments within the precipitate increases due to the larger modulus and the applied shear stress necessary to shear the precipitate decreases. Thus, the CRSS is attained right after precipitate shearing has started. 

Hence, the elastic mismatch between matrix and precipitate always leads to an increase in the CRSS (as compared with the homogeneous case) when the dislocation shears the precipitate, regardless of whether the precipitate shear modulus is larger or smaller than that of  the matrix. It should be noted that the dislocation will shear the precipitate without any hardening in the simulations presented in Fig. \ref{Fig:Curves_Shearable_NoFric_Hetero} if the matrix and precipitate have the same shear modulus.  The linear relationship between the CRSS and the elastic mismatch in the case of shearable precipitates is a result of the elastic effects that control the dislocation/precipitate interactions. In the case $G^P < G^M$, precipitate shearing is controlled by the Peach-Koehler force on the dislocation segments inside the precipitate, which is proportional to $G^P/G^M$. In the case $G^P > G^M$, the CRSS is determined by the  repulsive image stresses on the dislocation line as it approaches the precipitate. These image stresses are also proportional to $G^P/G^M$.

\begin{figure}[t]
\centering
	\includegraphics[width=0.49\textwidth]{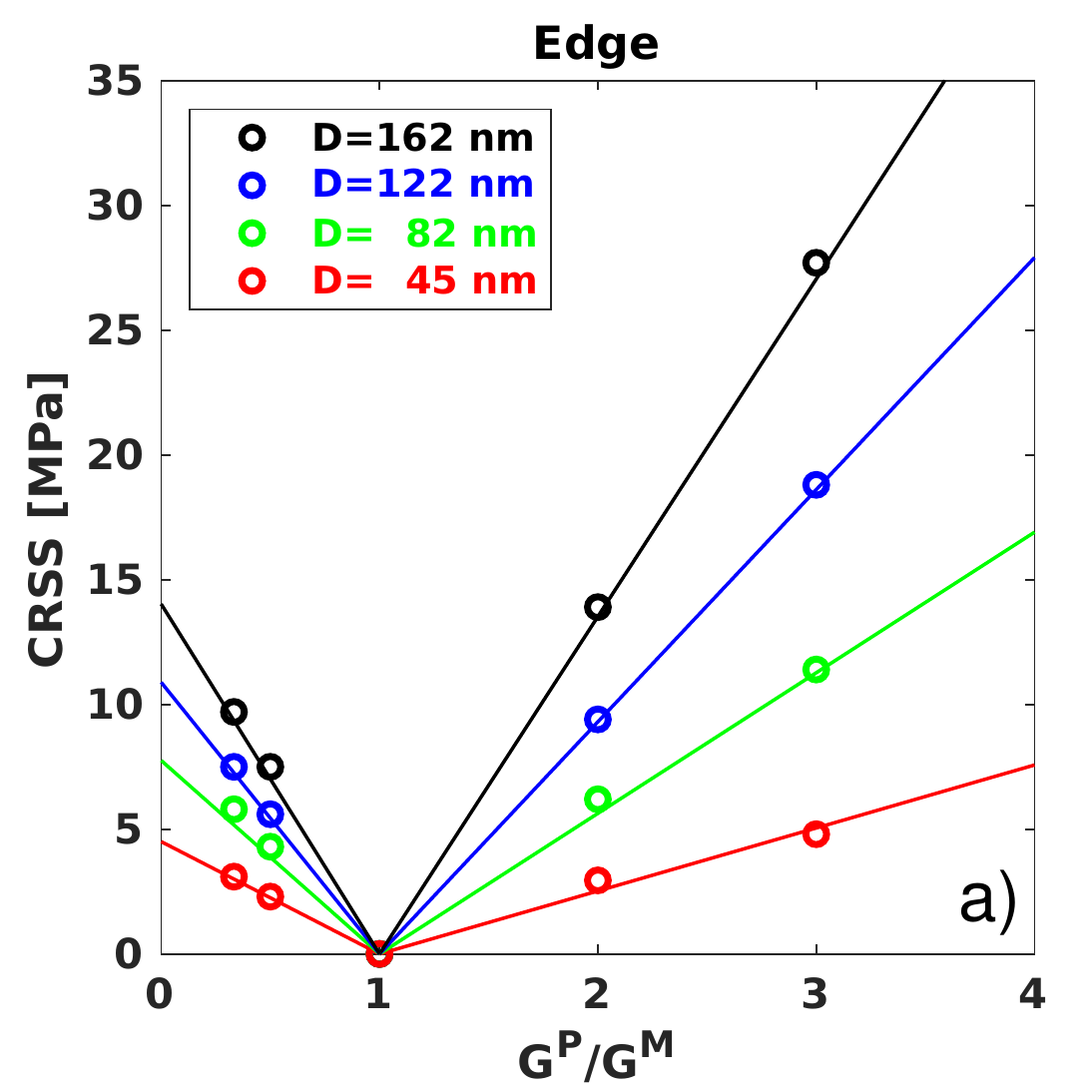}
	\includegraphics[width=0.49\textwidth]{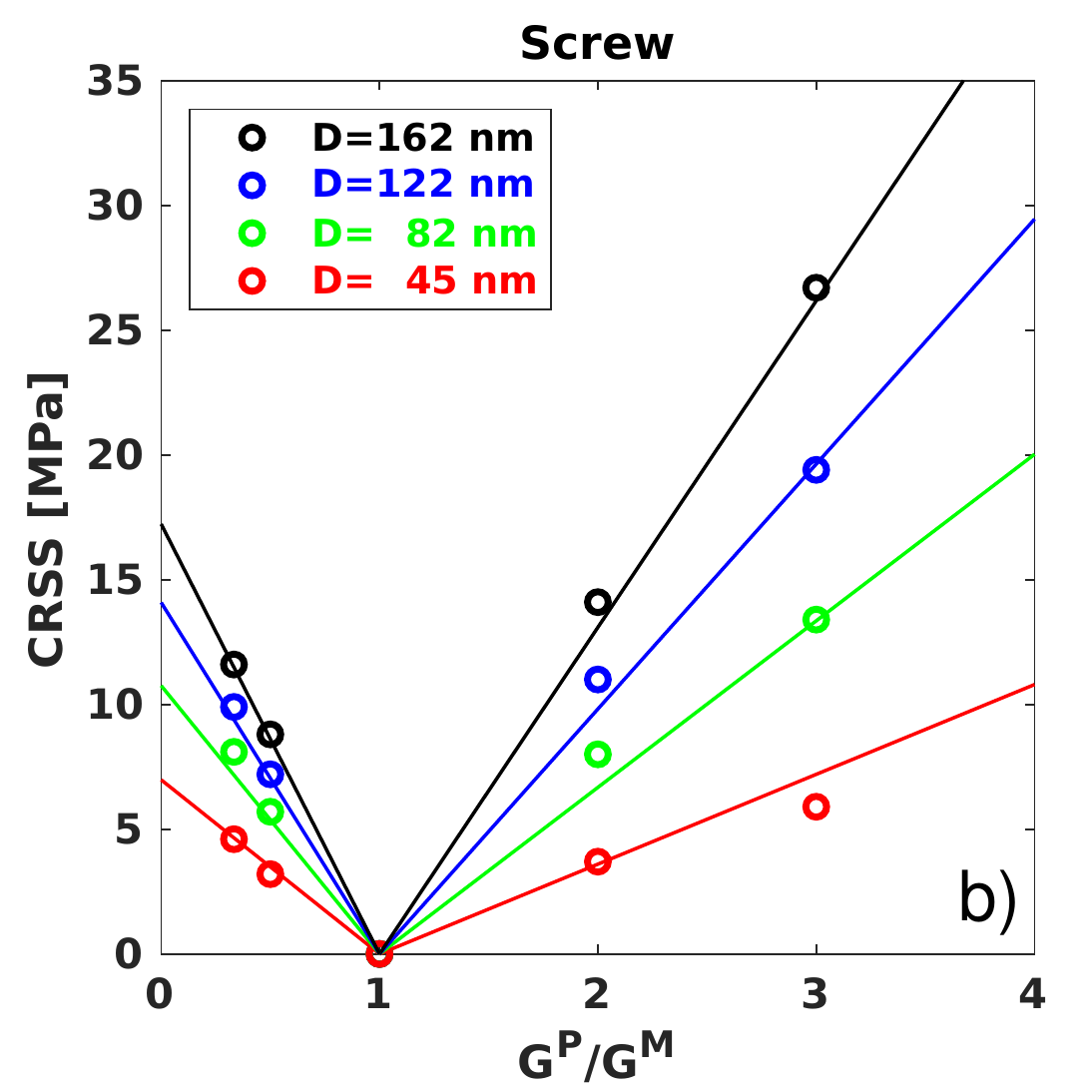}
	\caption{CRSS to shear precipitates without friction stress. The results obtained from DD simulations (circles) are compared with the predictions of eq. \eqref{Eq:ShearableHeteroNoFric} (solid lines) for different precipitate sizes and elastic mismatch. (a) Edge dislocation. (b) Screw dislocation. }
	\label{Fig:Shearable_NoFric}
\end{figure}

DD simulations were performed assuming a regular square array of precipitates stiffer and more compliant than the matrix and with different diameters in the slip plane. The strain rate was reduced to $10^4 \mathrm{s}^{-1}$ in the simulations with lower values of the CRSS (small precipitates without friction) in order to avoid strain rate effects. In addition, it was checked again that the influence of the strain rate ($5 \cdot 10^4 \mathrm{ s}^{-1}$) was negligible in the simulations with larger precipitates or higher friction stresses.  The CRSSs obtained from the DD simulations are plotted in Figs. \ref{Fig:Shearable_NoFric}a) and b) for edge and screw dislocations, respectively, as a function of $G^P/G^M$ for four different precipitate diameters. The dependence of  the CRSS with the elastic mismatch, $\Delta G = G^P- G^M$, the precipitate diameter $D$ and the centre-to-centre distance between precipitates $L_{ctc}$ could be expressed as

\begin{equation}
\tau_c^{sh}=K_1|\Delta G|\left(\frac{D}{L_{ctc}}\right)^{K_2}
\label{Eq:ShearableHeteroNoFric}
\end{equation}

\noindent where $K_1$ and $K_2$ are constants that depend on the dislocation character and on whether the precipitate is stiffer or more compliant than the matrix. These parameters are given in table \ref{Tab:Constants_K1_K2_K3}. The  predictions of the CRSS according to eq. \eqref{Eq:ShearableHeteroNoFric} are plotted in Figs. \ref{Fig:Shearable_NoFric}a and b for edge and screw dislocations, respectively, as a function of $G^P/G^M$ for four different precipitate diameters and they are in good agreement with the DD results.

\begin{center}
\begin{table}
    \begin{subtable}{0.49\textwidth}
        \centering
        \begin{tabular}{c c c c}
         & $K_1$ & $K_2$ & $K_3$ \\
        \toprule
 Edge & $1.7\cdot 10^{-3}$ & 1.3 & 0.5\\
 Screw & $1.25\cdot 10^{-3}$ & 1 & 1\\
 \bottomrule
       \end{tabular}
       \vspace*{3mm}
       \caption{$G^P>G^M$}
       \label{Tab:Constants_K1_K2_K3_>}
    \end{subtable}
    \hfill
    \begin{subtable}{0.49\textwidth}
        \centering
        \begin{tabular}{c c c c}
         & $K_1$ & $K_2$ & $K_3$ \\
        \hline
 Edge & $1.25\cdot 10^{-3}$ & 0.9 & 0.4 \\
 Screw & $1.25\cdot 10^{-3}$ & 0.7 & 0.4 \\
 \bottomrule
       \end{tabular}
       \vspace*{3mm}
       \caption{$G^P<G^M$}
       \label{Tab:Constants_K1_K2_K3_<}
     \end{subtable}
     \caption{Values of the constants $K_1$, $K_2$ and $K_3$ in eq. \eqref{Eq:ShearableHeteroNoFric} and \eqref{Eq:ShearableHeteroFric} for edge and screw dislocations.}
     \label{Tab:Constants_K1_K2_K3}
\end{table}
\end{center}

\begin{figure}
\centering
	\includegraphics[width=0.49\textwidth]{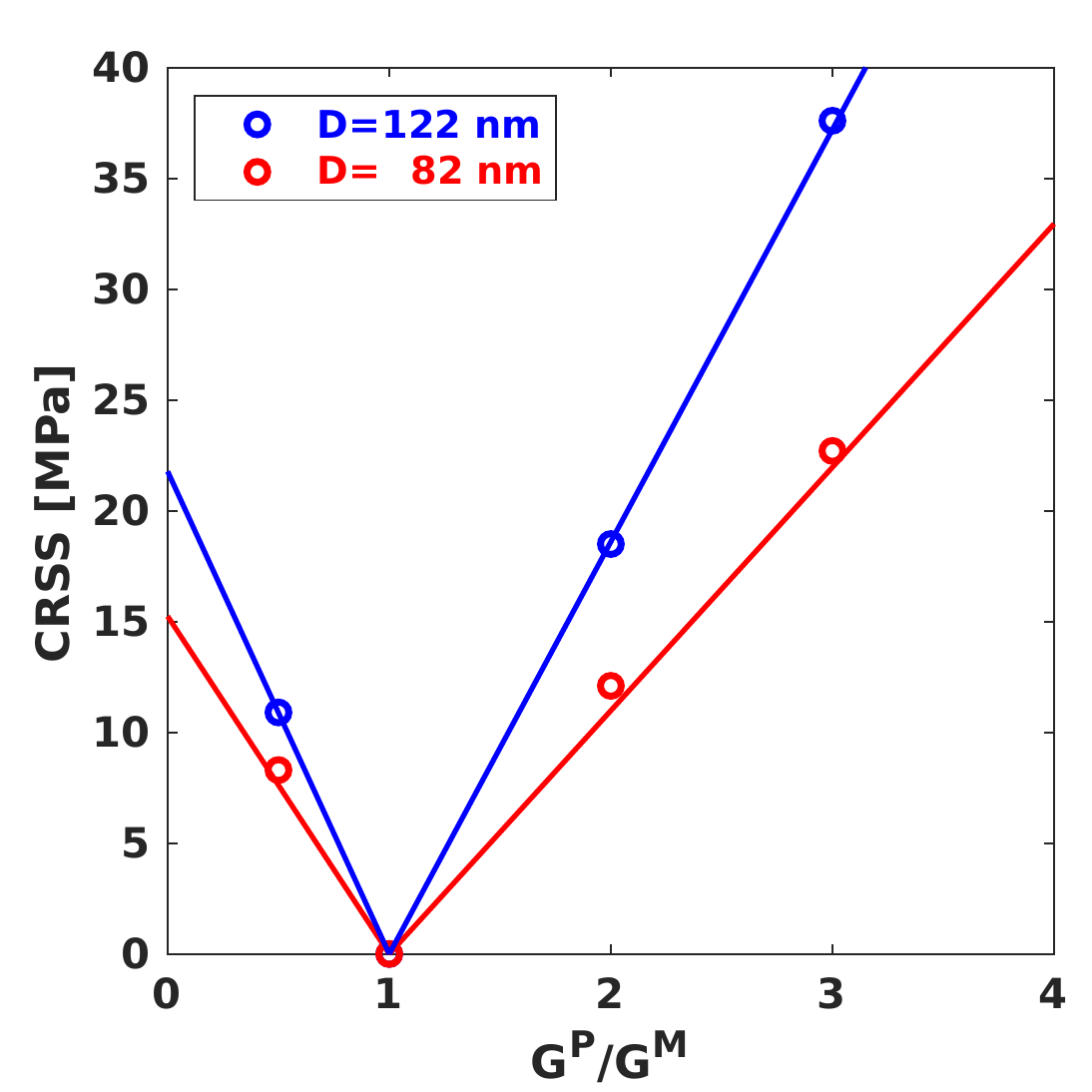}
	\caption{CRSS to shear precipitates without friction stress for a matrix with $G^M$ = 52.35 GPa. The results obtained from DD simulations (circles) are compared with the predictions of eq. \eqref{Eq:ShearableHeteroNoFric} (solid lines) for different precipitate sizes and elastic mismatch. The initial dislocation character was edge.}
	\label{Fig:LargeG}
\end{figure}

In order to check that the constants $K_1$ and $K_2$ were independent of the particular elastic constants used so far for the matrix, new simulations were performed in which the value of $G^M$ was increased by a factor of two ($G^M$ = 52.35 GPa), and the values of $G^P$ were increased accordingly. Simulations were carried out with two different precipitate sizes and an initial edge dislocation. The CRSS obtained from the  DD simulations and the predictions of eq. \eqref{Eq:ShearableHeteroNoFric} are plotted in Fig. \ref{Fig:LargeG}. The agreement of the new DD simulations with eq. \eqref{Eq:ShearableHeteroNoFric} was also very good, regardless of the shear modulus of the matrix and, thus,  the values of $K_1$ and $K_2$ in eq. \eqref{Eq:ShearableHeteroNoFric} can be considered independent of the elastic constants.

A general model of precipitate shearing by dislocations has to include the effects of  the friction stress and of the image stresses due to elastic heterogeneity. However, direct superposition of the models for the friction stress, eq. \eqref{Eq:ShearableAnalytical}, and for the image stresses, eq. \eqref{Eq:ShearableHeteroNoFric}, is not valid because the elastic mismatch modifies the influence of the former mechanism. Therefore, a prefactor depending on the shear modulus of the matrix and the precipitate, that modulates the effect of the friction stress, is required to capture the complex shearing of the precipitate in presence of elastic mismatch. This objective can be accomplished with the following expression

\begin{equation}
\tau_c^{sh}=K_1|\Delta G|\left(\frac{D}{L_{ctc}}\right)^{K_2}+\left(\frac{G^M}{G^P}\right)^{K_3}\tau^P\frac{D}{L_{ctc}}
\label{Eq:ShearableHeteroFric}
\end{equation}

\noindent  where the first term accounts for the effect of the image stresses on the CRSS (eq. \eqref{Eq:ShearableHeteroNoFric}) and the second term is identical to eq. \eqref{Eq:ShearableAnalytical} but includes a prefactor that depends on the ratio between matrix and precipitate elastic constants. The constants $K_1$ and $K_2$ in eq. \eqref{Eq:ShearableHeteroFric} are those from eq. \eqref{Eq:ShearableHeteroNoFric} (that can be found in table \ref{Tab:Constants_K1_K2_K3}), and the constant $K_3$ modulates the influence of the elastic mismatch on the friction stress.  $K_3$  depends on whether the precipitate is stiffer or more compliant than the matrix and on the dislocation character and the values for each case can also be found in table \ref{Tab:Constants_K1_K2_K3}. $K_3$ was determined for each case by fitting the predictions of the CRSS obtained with eq. \eqref{Eq:ShearableHeteroFric} to the results obtained with a regular square array of spherical precipitates with different diameters $D$, center-to-center spacing $L_{ctc}$, elastic mismatch $\Delta G$ and friction stress $\tau^P$ for either edge or screw dislocations. Eq. \eqref{Eq:ShearableHeteroFric} is, thus, a line tension model of precipitate shearing by dislocations that recovers the particular models when the medium is homogeneous ($G^P=G^M, \Delta G = 0$, eq. \eqref{Eq:ShearableAnalytical}) or the friction  stress in the precipitate is null ($\tau^P=0$, eq. \eqref{Eq:ShearableHeteroNoFric}). The validation of the model with $\tau^P \neq 0$ is presented below.

\section{Generalized line tension}

\begin{figure}
\centering
	\includegraphics[width=0.49\textwidth]{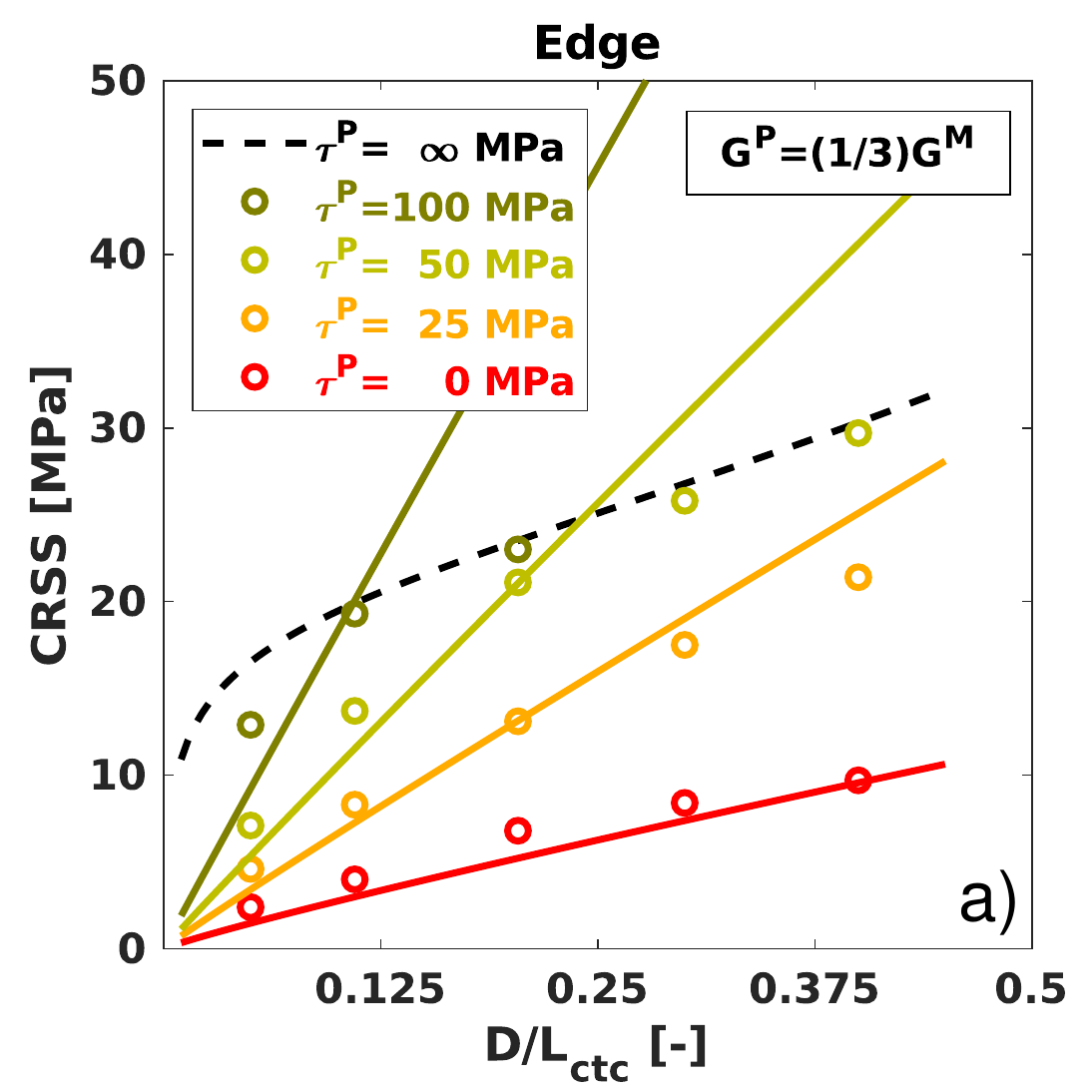}
	\includegraphics[width=0.49\textwidth]{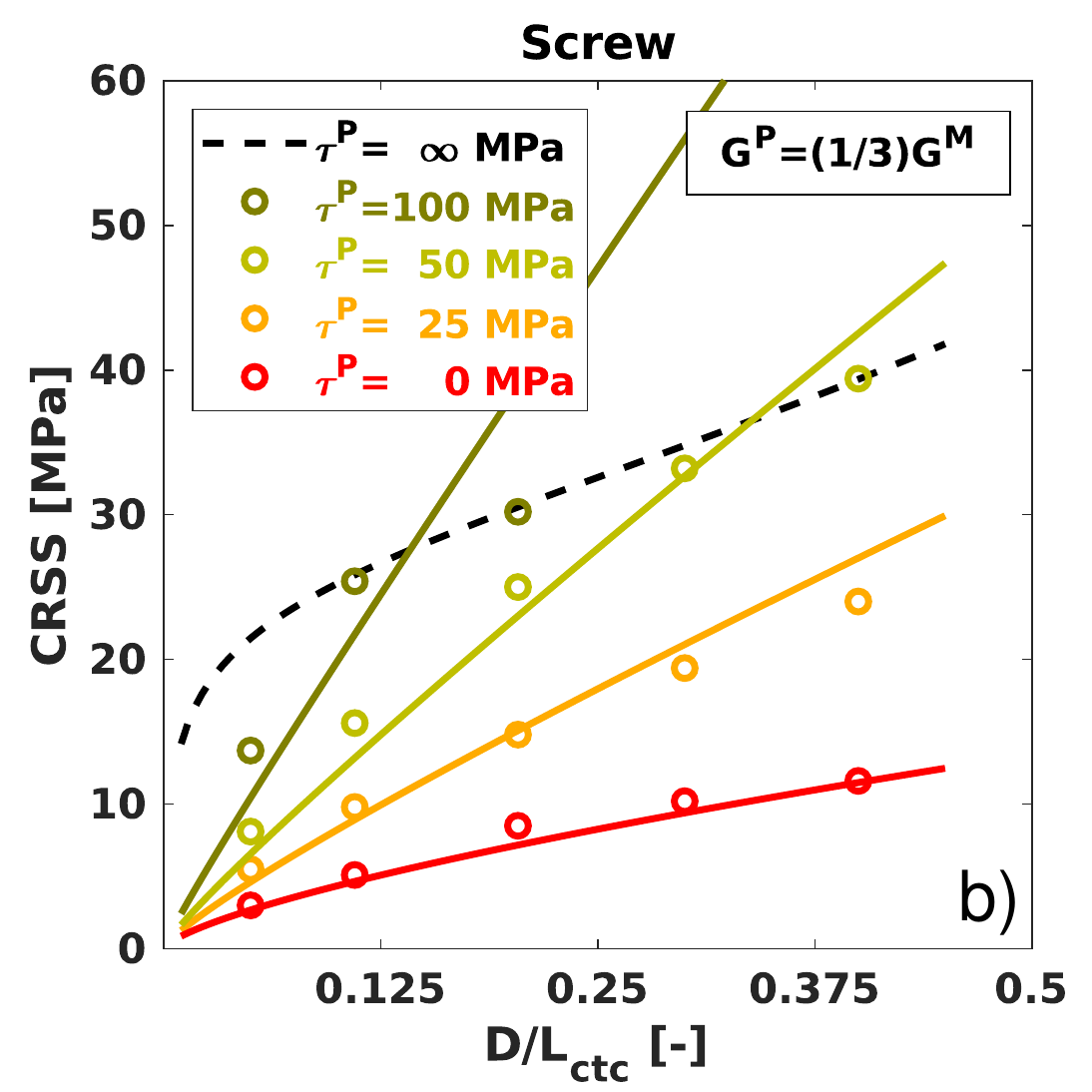}
	\caption{Comparison of the predictions of the CRSS with the generalized model, eq. \eqref{Eq:GeneralModel}, with the results of DD simulations. (a) Edge dislocation. (b) Screw dislocation. The dashed black line represents the model predictions for impenetrable precipitates while the solid lines stand for the results for shearable precipitates.  The DD results are plotted as open circles. $G^P/G^M = 1/3$.}
	\label{Fig:Shearable_0.3xC}
\end{figure}

\begin{figure}
\centering
	\includegraphics[width=0.49\textwidth]{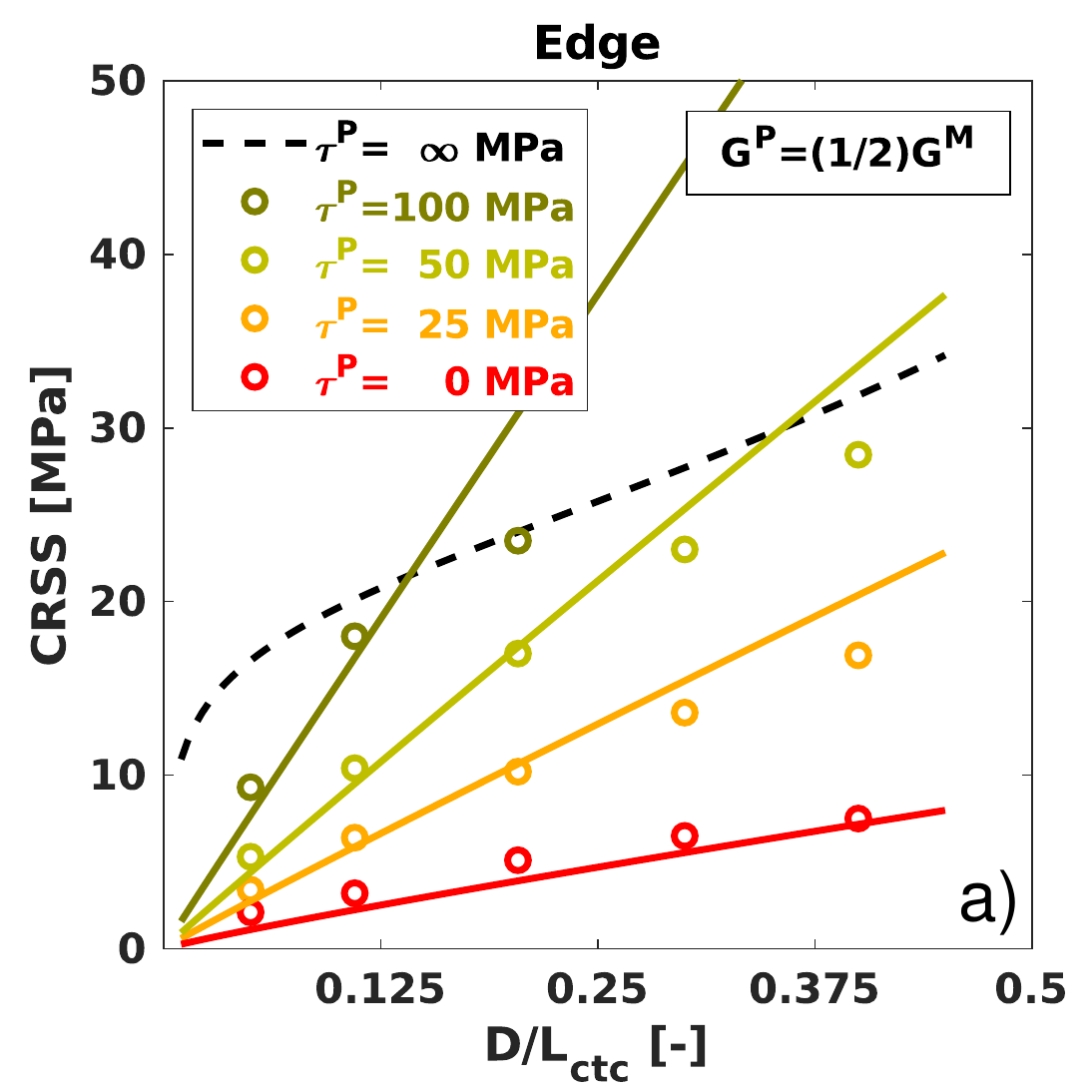}
	\includegraphics[width=0.49\textwidth]{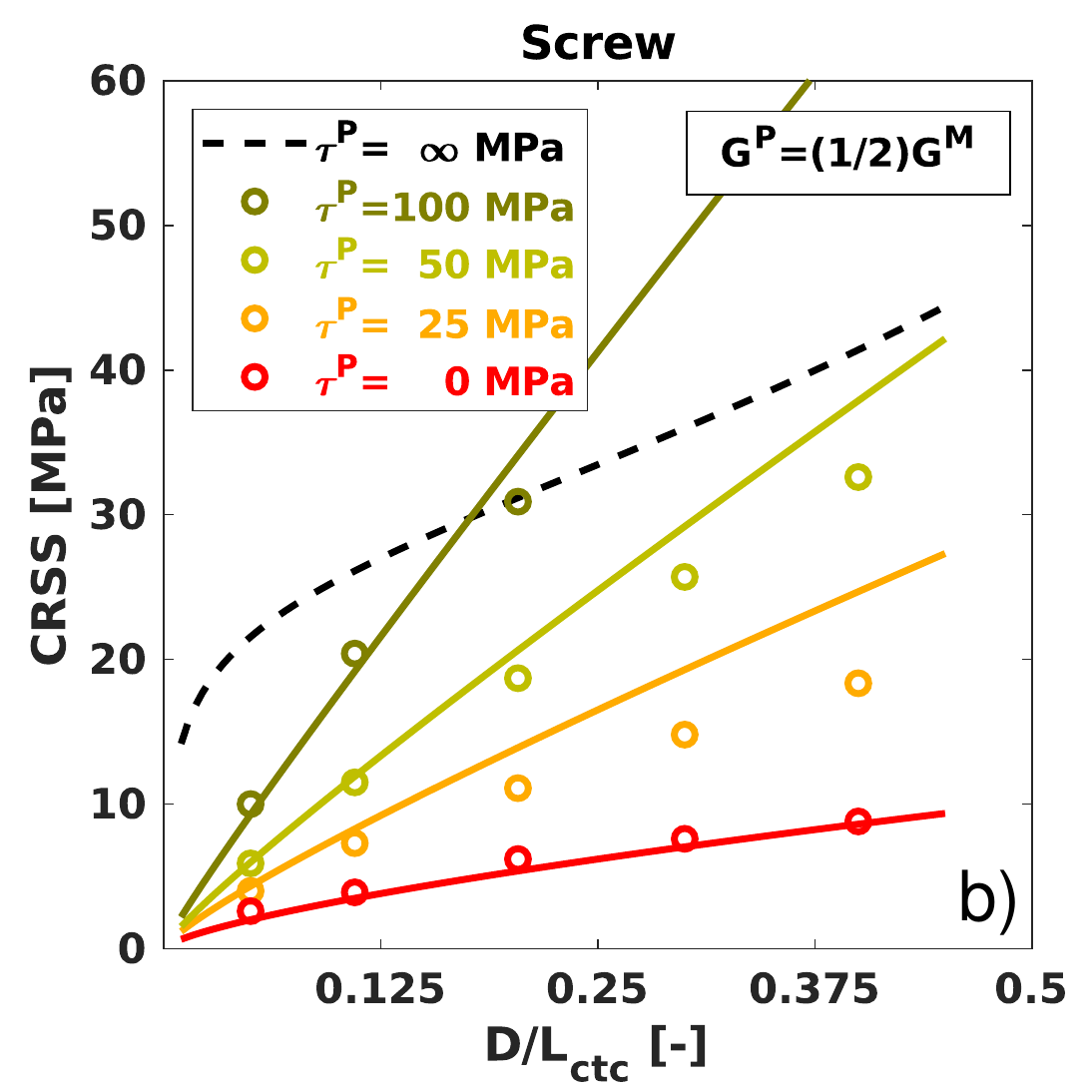}
	\caption{Comparison of the predictions of the CRSS with the generalized model, eq. \eqref{Eq:GeneralModel}, with the results of DD simulations. (a) Edge dislocation. (b) Screw dislocation. The dashed black line represents the model predictions for impenetrable precipitates while the solid lines stand for the results for shearable precipitates.  The DD results are plotted as open circles. $G^P/G^M = 1/2$.}
		\label{Fig:Shearable_0.5xC}
\end{figure}

\begin{figure}
\centering
	\includegraphics[width=0.49\textwidth]{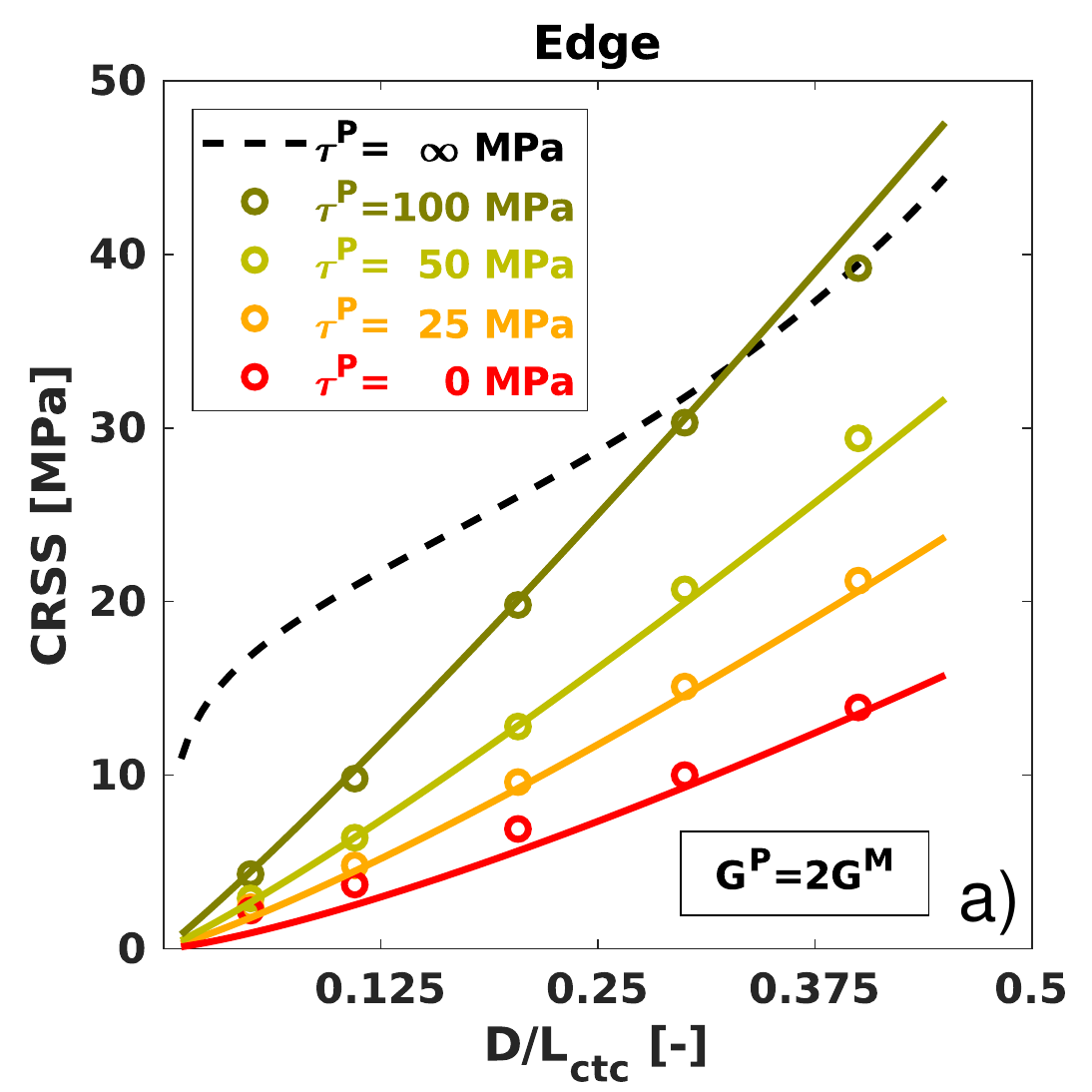}
	\includegraphics[width=0.49\textwidth]{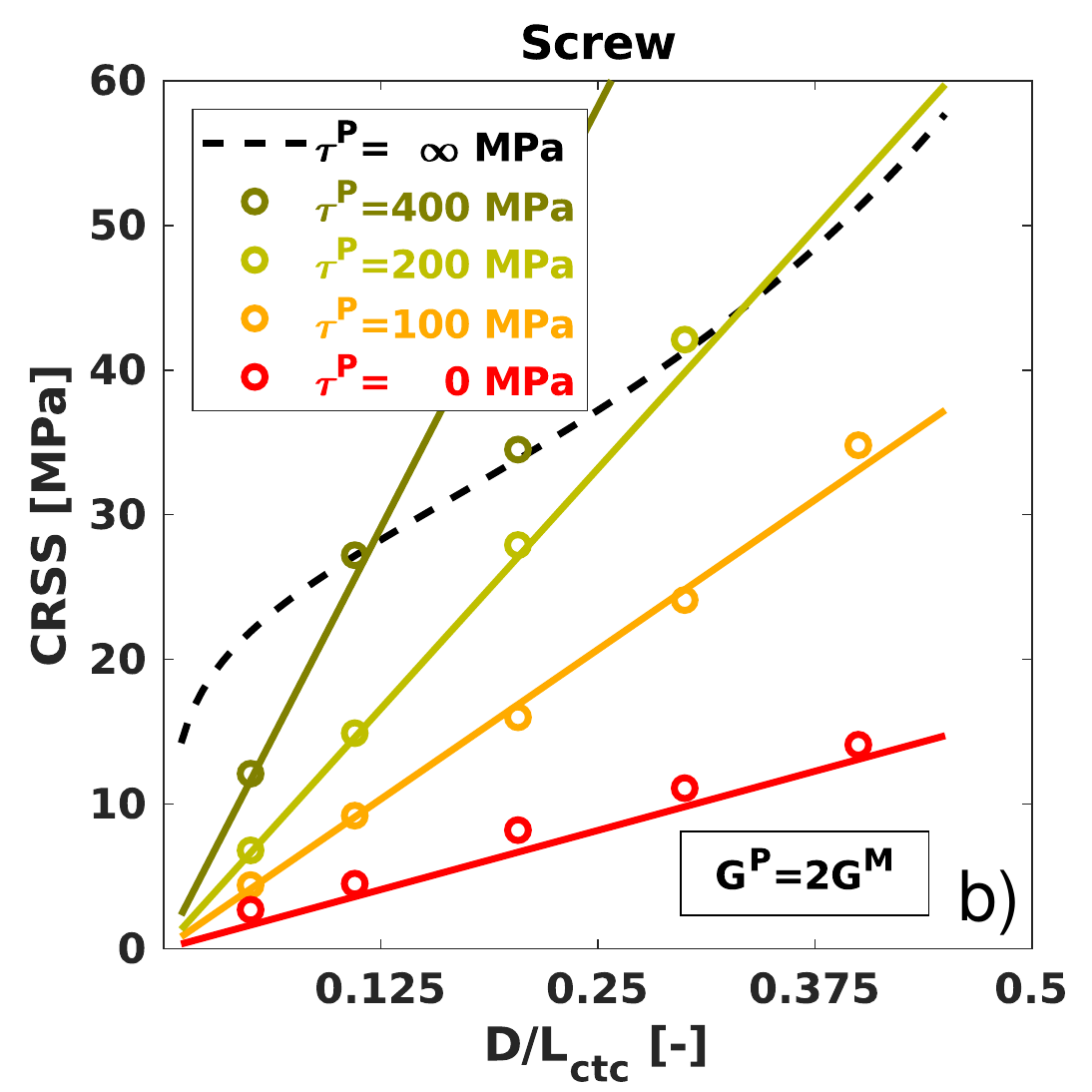}
	\caption{Comparison of the predictions of the CRSS with the generalized model, eq. \eqref{Eq:GeneralModel}, with the results of DD simulations. (a) Edge dislocation. (b) Screw dislocation. The dashed black line represents the model predictions for impenetrable precipitates while the solid lines stand for the results for shearable precipitates.  The DD results are plotted as open circles. $G^P/G^M = 2$.}
		\label{Fig:Shearable_2xC}
\end{figure}

\begin{figure}
\centering
	\includegraphics[width=0.49\textwidth]{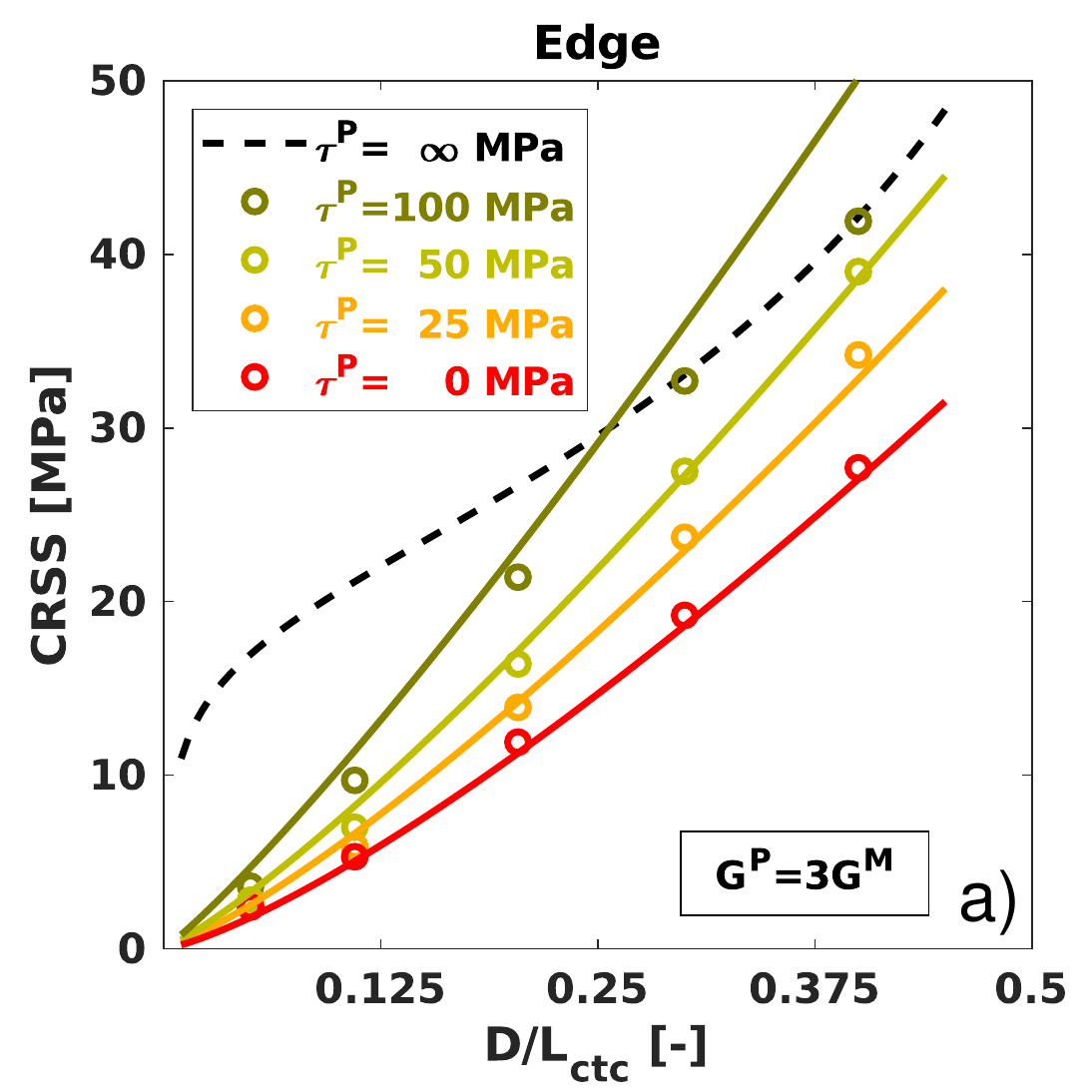}
	\includegraphics[width=0.49\textwidth]{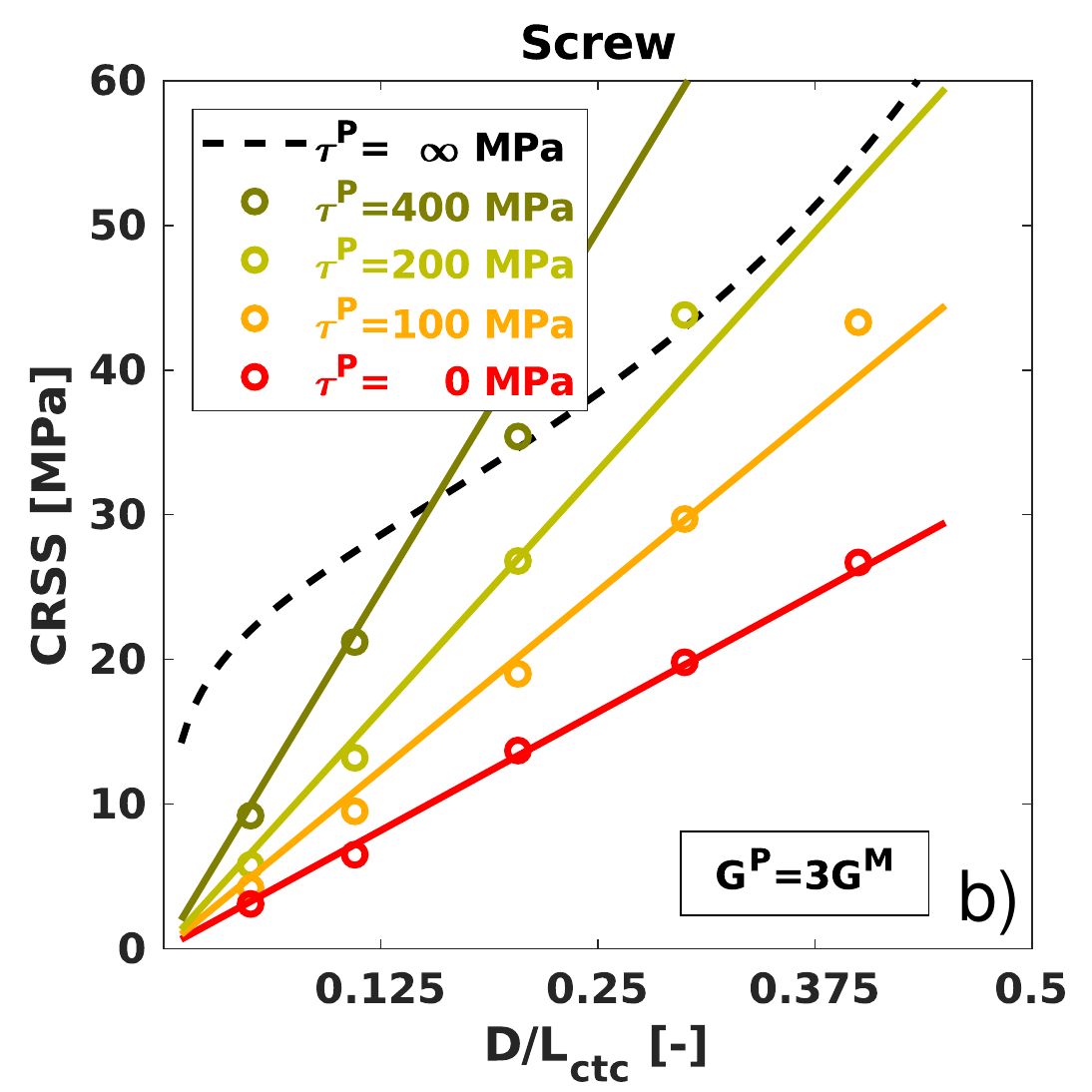}
	\caption{Comparison of the predictions of the CRSS with the generalized model, eq. \eqref{Eq:GeneralModel}, with the results of DD simulations. (a) Edge dislocation. (b) Screw dislocation. The dashed black line represents the model predictions for impenetrable precipitates while the solid lines stand for the results for shearable precipitates.  The DD results are plotted as open circles. $G^P/G^M = 3$.}
	\label{Fig:Shearable_3xC}
\end{figure}

The line tension models developed above account for the by-pass of the precipitates by either shearing or the formation of an Orowan loop. When both mechanisms are possible (because the precipitates can be sheared by dislocations), the CRSS will be given by the lowest value necessary to overcome the precipitate as given by the line tension model for impenetrable precipitates, eq. \eqref{Eq:BKS_Heterogeneous}, and the line tension model for shearable precipitates, eq. \eqref{Eq:ShearableHeteroFric}:

\begin{equation}
\tau_c=\min(\tau_c^{imp},\tau_c^{sh})
\label{Eq:GeneralModel}
\end{equation}

\noindent and the equations of each of the models have been summarized in table \ref{Tab:Summary} for the sake of clarity. The CRSSs obtained with this generalized line tension model are compared with those determined by means of DD in Figs. \ref{Fig:Shearable_0.3xC}-\ref{Fig:Shearable_3xC}. Each figure includes the results obtained with either edge or screw dislocations for one magnitude of the elastic mismatch ($G^P/G^M$ = 1/3, 1/2, 2 and 3) and the CRSS is plotted as a function of $D/L_{ctc}$ for different values of the friction stress of the precipitate. Obviously, $\tau^P = \infty$ stands for impenetrable precipitates. The impenetrable regime is represented by a dashed black line, corresponding to eq. \eqref{Eq:BKS_Heterogeneous}, while the generalized line tension model predictions for shearable precipitates, eq. \eqref{Eq:ShearableHeteroFric}, are shown by solid lines whose colour depends on the friction stress of the precipitate. The agreement between DD simulations and the generalized line tension model is very good. Furthermore, the generalized model is able to capture the transition between the impenetrable and shearable regimes.

\begin{center} 
\begin{table}[t]
\begin{center}
\begin{tabular}{c|c|c}

 & \textbf{Shearable} & \textbf{Impenetrable} \\
\hline
& & \\
 $G^M=G^P$ & $\tau_c^{sh}=\tau^P\frac{D}{L_{ctc}}$ & $\tau_c^{imp}=\frac{G^Mb}{L}A\left[\ln\left(\frac{\bar{D}}{b}\right)+B \right]$ \\
 & & \\
 $G^M \neq G^P$ & $\tau_c^{sh}=K_1|\Delta G|\left(\frac{D}{L_{ctc}}\right)^{K_2}+\left(\frac{G^M}{G^P}\right)^{K_3}\tau^P\frac{D}{L_{ctc}}$ & $\tau_c^{imp}=\frac{G^Mb}{L_{eff}}A\left[\ln\left(\frac{\bar{D}}{b}\right)+B \right]$\\
 & & \\

\hline
\end{tabular}
\end{center}
\caption{Summary of the expressions of the line tension models to predict the CRSS of a regular square array of spherical precipitates. See text for details.}
\label{Tab:Summary}
\end{table}
\end{center}

\section{Strengthening of random precipitate distributions}

The generalized line tension model developed above provides accurate results for the CRSS in the case of a regular square array of spherical precipitates in the slip plane. Nevertheless, precipitate distributions in metallic alloys are usually random and it is interesting to extend the results of the generalized line tension model to this scenario. A  simple approach to extend the generalized line tension model to random precipitate distributions  is based on the modification of the geometrical parameters of the model. 

Regarding the precipitate diameter $D$, it should be noticed that the slip plane may intersect the precipitate at any height in a random distribution. Thus, the "planar diameter" (understood as the mean diameter of the circular intersection between the slip plane and the spherical precipitate) may range from 0 (no intersection) to $D$ (intersection at the centre of the sphere). For the case of a random distribution of spherical precipitates of constant diameter $D$, the mean planar diameter is given by \citep{Nembach1997}:

\begin{equation}
\langle D\rangle=\frac{\pi}{4}D.
\label{Eq:MeanPlanarD}
\end{equation}

The average inter-precipitate distance $\langle L\rangle$ can be computed as:

\begin{equation}
\langle L\rangle=\langle L_{ctc}\rangle-\langle D\rangle
\label{Eq:LRandDist}
\end{equation}

\noindent where $\langle L_{ctc}\rangle$ is the average center-to-center distance between precipitates, which is equal to the square lattice spacing in the case of a regular square array. The average $\langle L_{ctc}\rangle$  in the case of a random distribution of spherical precipitates with constant diameter $D$ is given by \citep{Nembach1997}:

\begin{equation}
\langle L_{ctc}\rangle=\frac{D}{2}\sqrt{\frac{2\pi}{3f}}
\label{Eq:Lctc}
\end{equation}

\noindent where $f$ stands for the precipitate volume fraction.

Finally, the average effective distance, $\langle L_{eff}\rangle$ is given by

\begin{equation}
\langle L_{eff}\rangle=\langle L\rangle \left(1-\frac{\langle D\rangle}{\langle L\rangle}\alpha\right)
\label{Eq:Leff_dist}
\end{equation}

\noindent while  $\overline{\langle{D}\rangle}$ stands for the average harmonic mean of $\langle{L}\rangle$ and the mean planar diameter $\langle{D}\rangle$, and it  is expressed as 

\begin{equation}
\overline{\langle{D}\rangle}=\left(\frac{1}{\langle D\rangle}+\frac{1}{\langle L\rangle}\right)^{-1}.
\end{equation}

Based on these "averaged" geometrical descriptors, new expressions for the CRSS in the case of either impenetrable or shearable precipitates can be obtained. In the case of impenetrable precipitates, the starting point is the BKS model for a random distribution of precipitates given by eq. \eqref{Eq:BKS_random}. This model can be easily modified to account for the elastic heterogeneity by replacing $\langle L\rangle$ by the effective inter-precipitate distance $\langle L_{eff}\rangle$  leading to

\begin{equation}
\tau_c^{BKS,rand}=A\frac{Gb}{\langle L_{eff}\rangle}  
\frac{\left[\ln\left(2\overline{\langle{D}\rangle}/b\right)\right]^{3/2}}{\left[\ln\left(\langle L_{eff}\rangle/b\right)\right]^{1/2}}
\label{Eq:BKS_random_heterogeneous}
\end{equation}

\noindent where $\bar{D}$ has been replaced by $\overline{\langle{D}\rangle}$ to account for the random precipitate distribution and the constant $A$ depends on the dislocation character. 

As in eq. \eqref{Eq:BKS_random}, the constant $A$ can be replaced by the geometric mean of $A$ for edge and screw dislocations, leading to
\begin{equation}
\tau_c^{BKS,rand}=\frac{1}{2\pi\sqrt{1-\nu}}\frac{Gb}{\langle L_{eff}\rangle}  
\frac{\left[\ln\left(2\overline{\langle{D}\rangle}/b\right)\right]^{3/2}}{\left[\ln\left(\langle L_{eff}\rangle/b\right)\right]^{1/2}}
\label{Eq:BKS_random_mean_heterogeneous}
\end{equation}

\noindent which gives the CRSS for mixture of edge and screw dislocations in the presence of a random distribution of  impenetrable, spherical precipitates.

In the case of shearable precipitates, the generalized line tension model (eq. \eqref{Eq:ShearableHeteroFric}) was simply modified by replacing $D$ and $L_{ctc}$  by the corresponding average planar diameter $\langle D \rangle$ and average center-to-center spacing $\langle L_{ctc}\rangle$, respectively. Thus, the generalized line tension model for a random arrangement of shearable precipitates is given by:

\begin{equation}
\tau_c^{sh,rand}=K_1|\Delta G|\left(\frac{\langle D\rangle}{\langle L_{ctc}\rangle}\right)^{K_2}+\left(\frac{G^M}{G^P}\right)^{K_3}\tau^P\frac{\langle D\rangle}{\langle  L_{ctc} \rangle}
\label{Eq:ShearableHeteroFric_Rand}
\end{equation}

\noindent where  $\langle D\rangle$ and $\langle L_{ctc}\rangle$ are given by eq. \eqref{Eq:MeanPlanarD} and \eqref{Eq:Lctc} respectively. The constants $K_1$, $K_2$ and $K_3$ in eq. \eqref{Eq:ShearableHeteroFric_Rand} are the same ones used for the regular square array of precipitates given in Table \ref{Tab:Constants_K1_K2_K3}. It should be noticed that they are different for edge and screw dislocations and also depend on whether the precipitate is stiffer than the matrix or viceversa.

\begin{figure}[t]
\centering
	\includegraphics[width=0.49\textwidth]{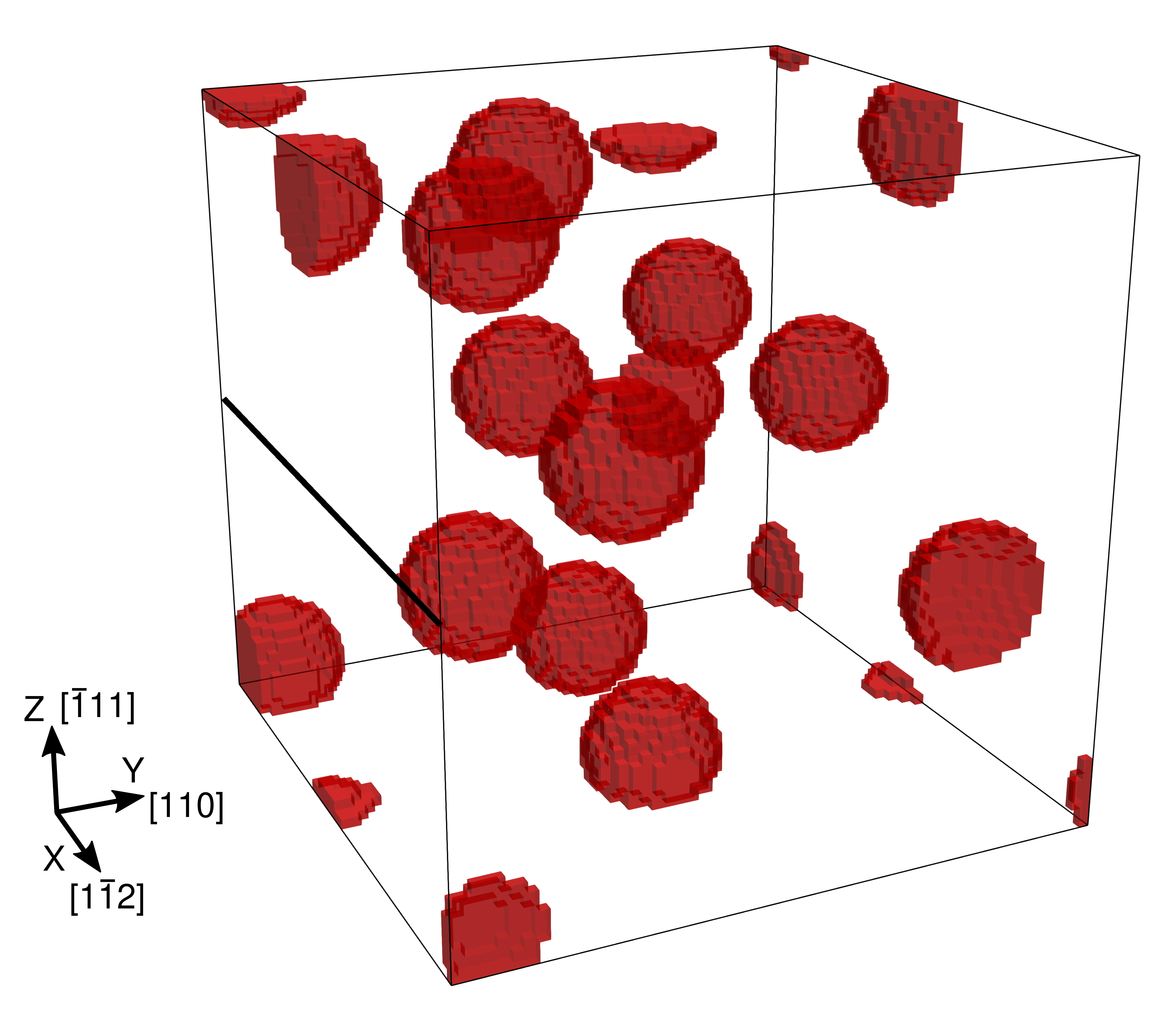}
	\caption{Cubic simulation domain containing 12 spherical precipitates of $D$ = 82 nm. The initial edge dislocation line is represented by a straight black line.}
	\label{Fig:Dist_spheres}
\end{figure}

In order to validate eqs. \eqref{Eq:BKS_random_heterogeneous} and \eqref{Eq:ShearableHeteroFric_Rand}, new DD simulations were carried out to determine the CRSS to overcome a random distribution of spherical precipitates. The periodic simulation domain of 405 x 405 x 405 nm$^3$ included 12 spherical precipitates randomly distributed. Two different precipitate diameters were used in the simulations ($D$ = 82 nm and 70 nm), leading to two different precipitate volume fractions ($f$ = 5.3\% and 3.2\%). 10 different precipitate realizations were generated for each precipitate diameter (or volume fraction) and simulations for each realization were performed assuming that the domain contained either an edge or screw straight dislocation segment. An example of the domain with 12 precipitates of 82 nm in diameter and an edge dislocation line is shown in Fig. \ref{Fig:Dist_spheres}. The orientations of the domains are equivalent to the ones in Fig. \ref{Fig:InitialSphere} and a shear strain rate of $5\cdot10^4 \rm{ s}^{-1}$ was applied to drive the dislocation motion. Three simulations were carried out for each precipitate distribution, assuming that the precipitates were either impenetrable to  dislocations or  could be sheared. In the latter case, two different values of the friction stress $\tau^P$ (either 50 MPa or 100 MPa) were used. Moreover, simulations were performed assuming that $G^P =G^M$ or $G^P =2G^M$.

\begin{figure}[!t]
\centering
	\includegraphics[width=0.49\textwidth]{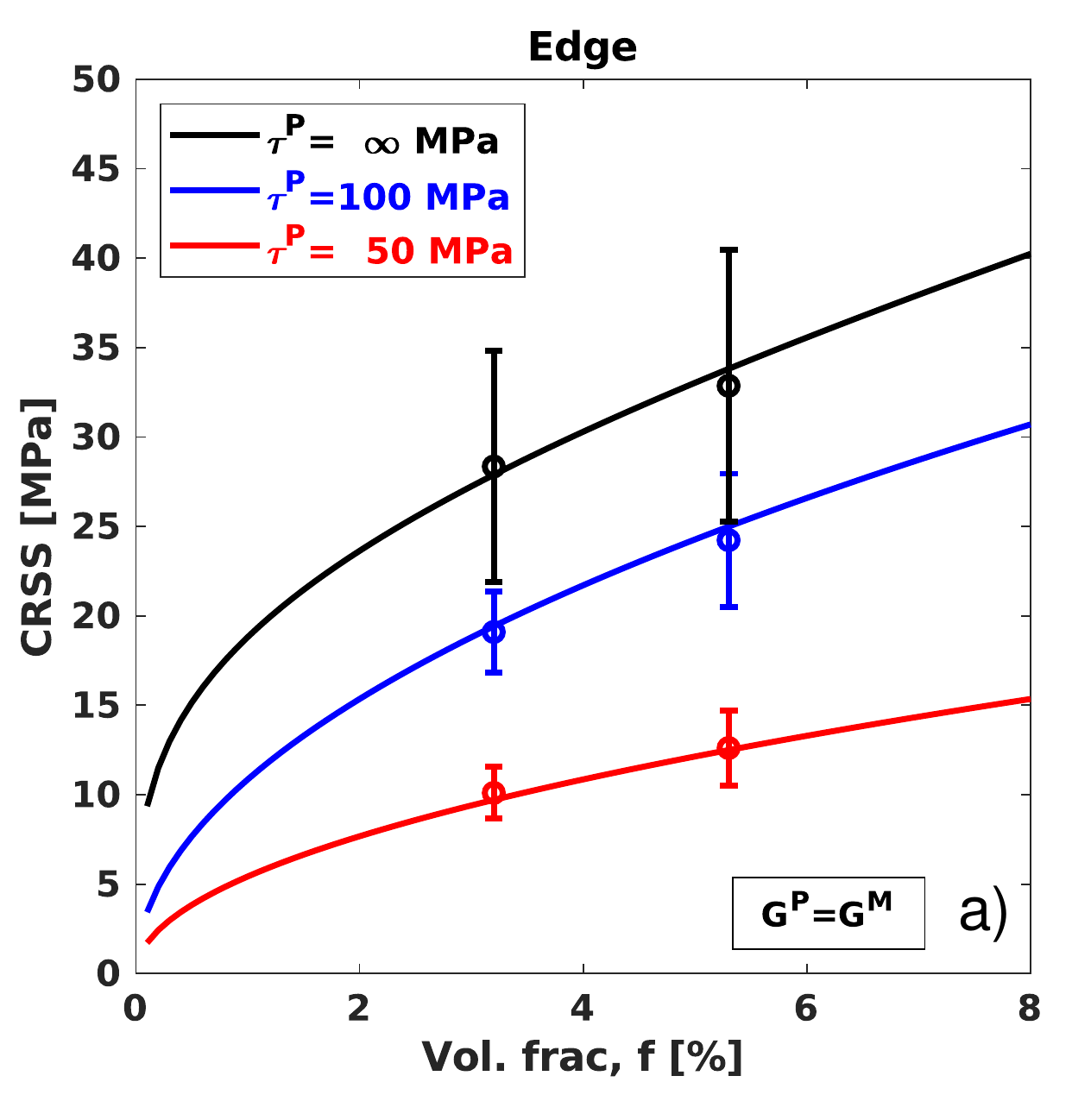}
	\includegraphics[width=0.49\textwidth]{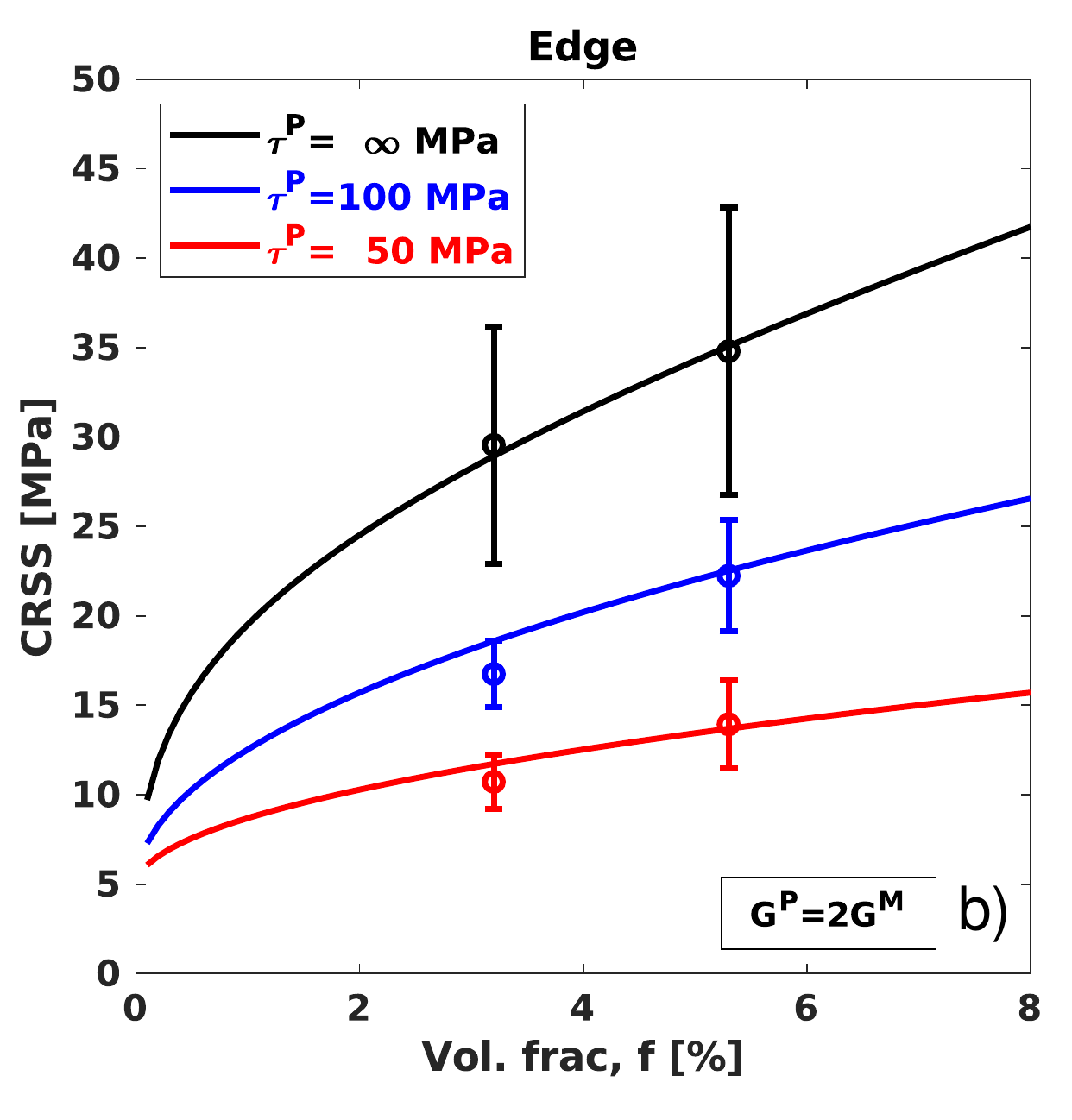}
	\caption{Comparison of the predictions of the CRSS with the generalized line tension model for edge dislocations and random precipitate distributions, eqs. \eqref{Eq:BKS_random_heterogeneous} and \eqref{Eq:ShearableHeteroFric_Rand}, with the results of DD simulations for different precipitate volume fractions.  The average values from the DD simulations are indicated by open circles, along with the error bars indicating the standard deviation. The model predictions are plotted as solid lines. (a) $G^P=G^M$ (b) $G^P=2G^M$}
	\label{Fig:N12_edge}
\end{figure}

\begin{figure}[!t]
\centering
	\includegraphics[width=0.49\textwidth]{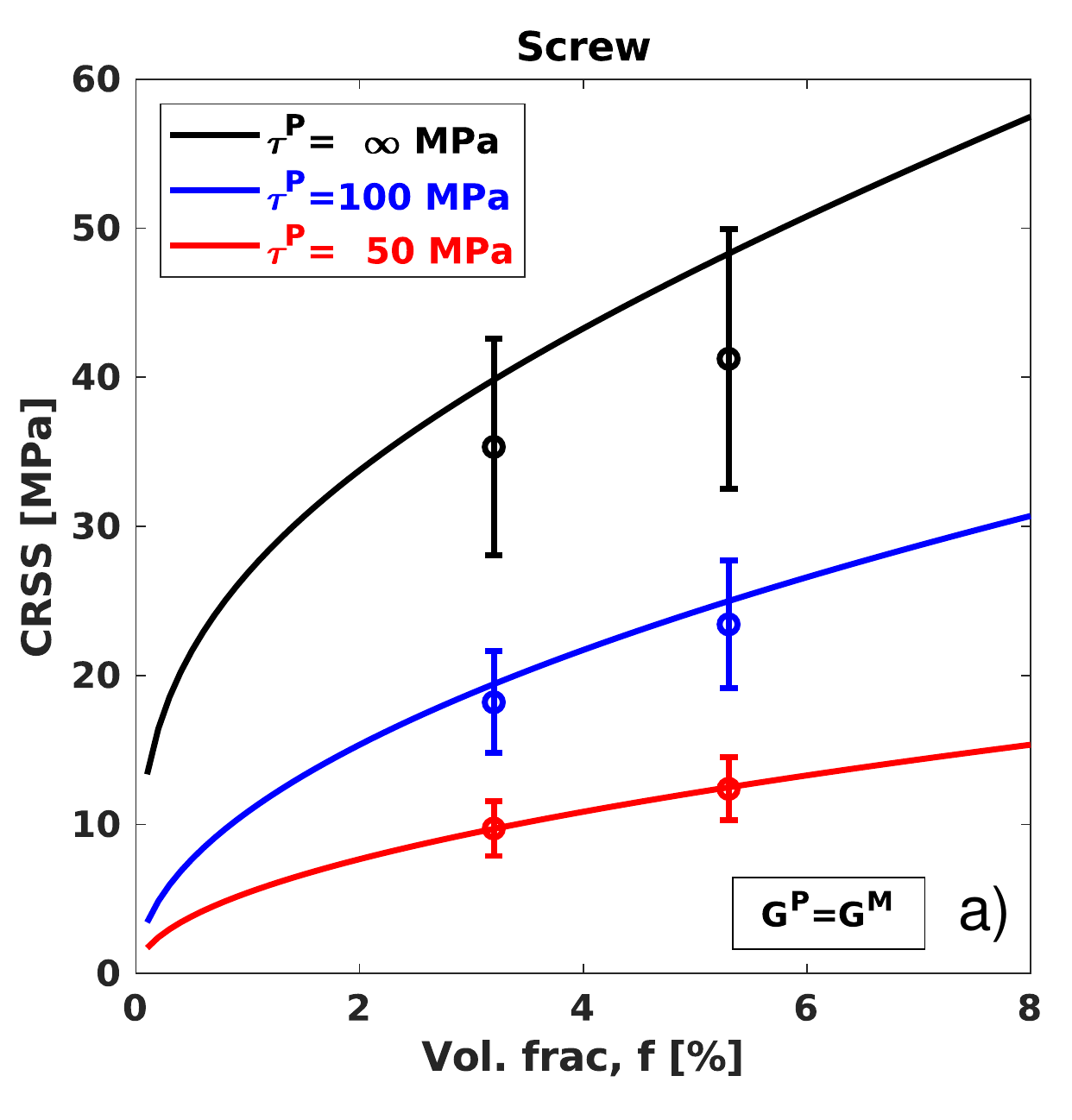}
	\includegraphics[width=0.49\textwidth]{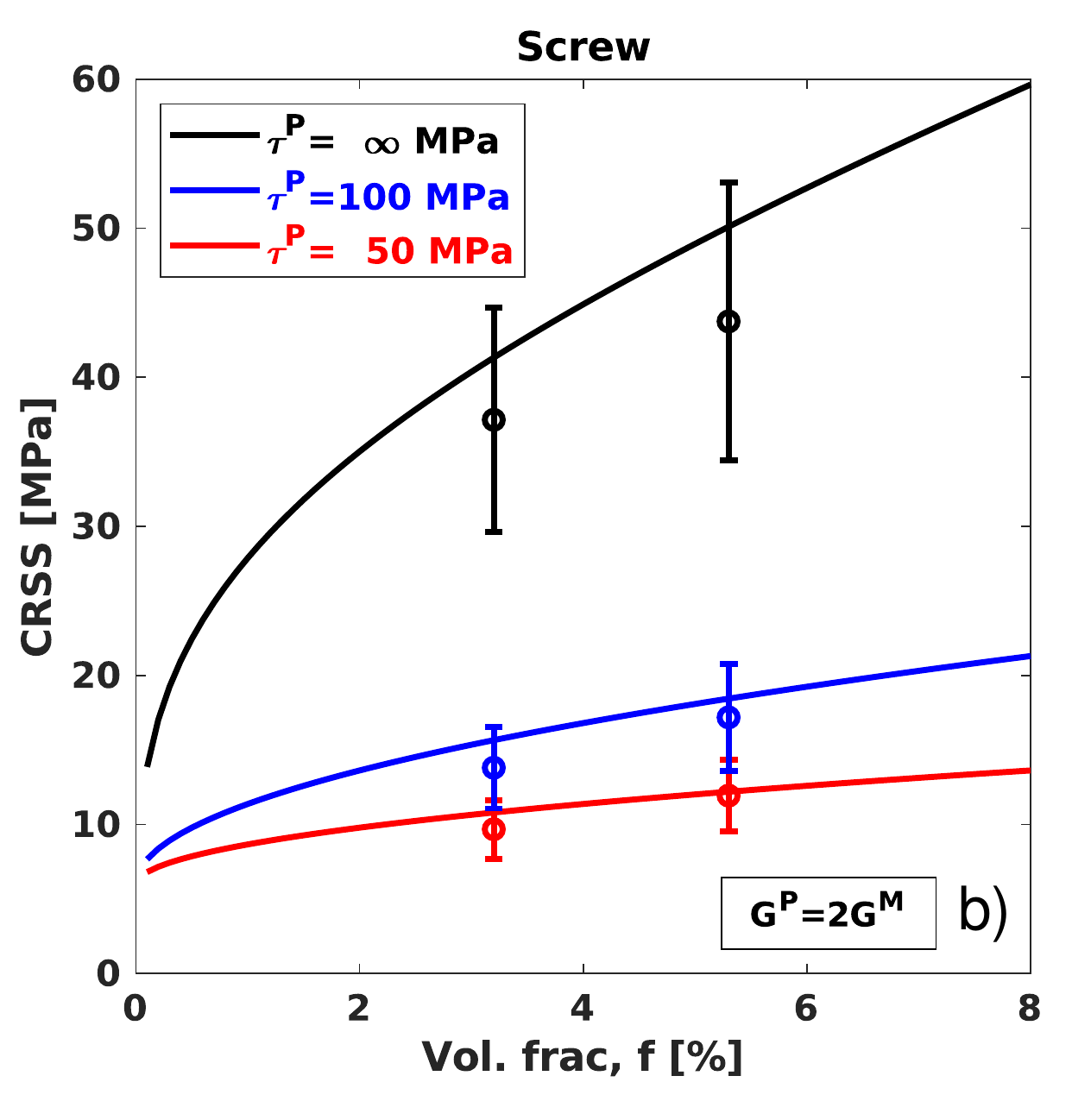}
	\caption{Comparison of the predictions of the CRSS with the generalized line tension model for screw dislocations and random precipitate distributions, eqs. \eqref{Eq:BKS_random_heterogeneous} and \eqref{Eq:ShearableHeteroFric_Rand}, with the results of DD simulations for different precipitate volume fractions.  The average values from the DD simulations are indicated by open circles, along with the error bars indicating the standard deviation. The model predictions are plotted as solid lines. (a) $G^P=G^M$ (b) $G^P=2G^M$}
	\label{Fig:N12_screw}
\end{figure}

The average values (and the corresponding standard deviations) of the CRSS obtained from 10 DD simulations with a initial edge dislocation are plotted in Figs. \ref{Fig:N12_edge}a and b  as a function of the precipitate volume fraction when $G^P =G^M$ or $G^P =2G^M$, respectively. Similarly, results corresponding to an initial screw dislocations are shown in Figs. \ref{Fig:N12_screw}a and b. The predictions of eqs. \eqref{Eq:BKS_random_heterogeneous} and \eqref{Eq:ShearableHeteroFric_Rand} for random distributions of impenetrable and shearable spherical precipitates, respectively, are plotted as solid lines in these figures. The agreement between the generalized line tension models and the results of the DD simulations is very good, particularly for shearable precipitates. In the case of impenetrable precipitates, the predictions of eq.  \eqref{Eq:BKS_random_heterogeneous} are also excellent for the case of edge dislocations and overestimate slightly the DD results for screw dislocations.

It is worth noting that the expression for the CRSS of shearable precipitates is a function of $\langle\bar{D}\rangle/ \langle L_{ctc}\rangle$, which -according to eqs. \eqref{Eq:MeanPlanarD} and \eqref{Eq:Lctc}- only depends on the precipitate volume fraction $f$ and not on other geometrical parameters. Thus, the CRSS for a random distribution of spherical shearable precipitates can be expressed as

\begin{equation}
\tau_c^{sh,rand}=K_1|\Delta G|\left(\sqrt{\frac{3\pi f}{8}}\right)^{K_2}+\left(\frac{G^M}{G^P}\right)^{K_3}\tau^P\sqrt{\frac{3\pi f}{8}}.
\label{Eq:ShearableHeteroFric_Rand_volFrac}
\end{equation}

\begin{figure}[t!]
\centering
	\includegraphics[width=0.59\textwidth]{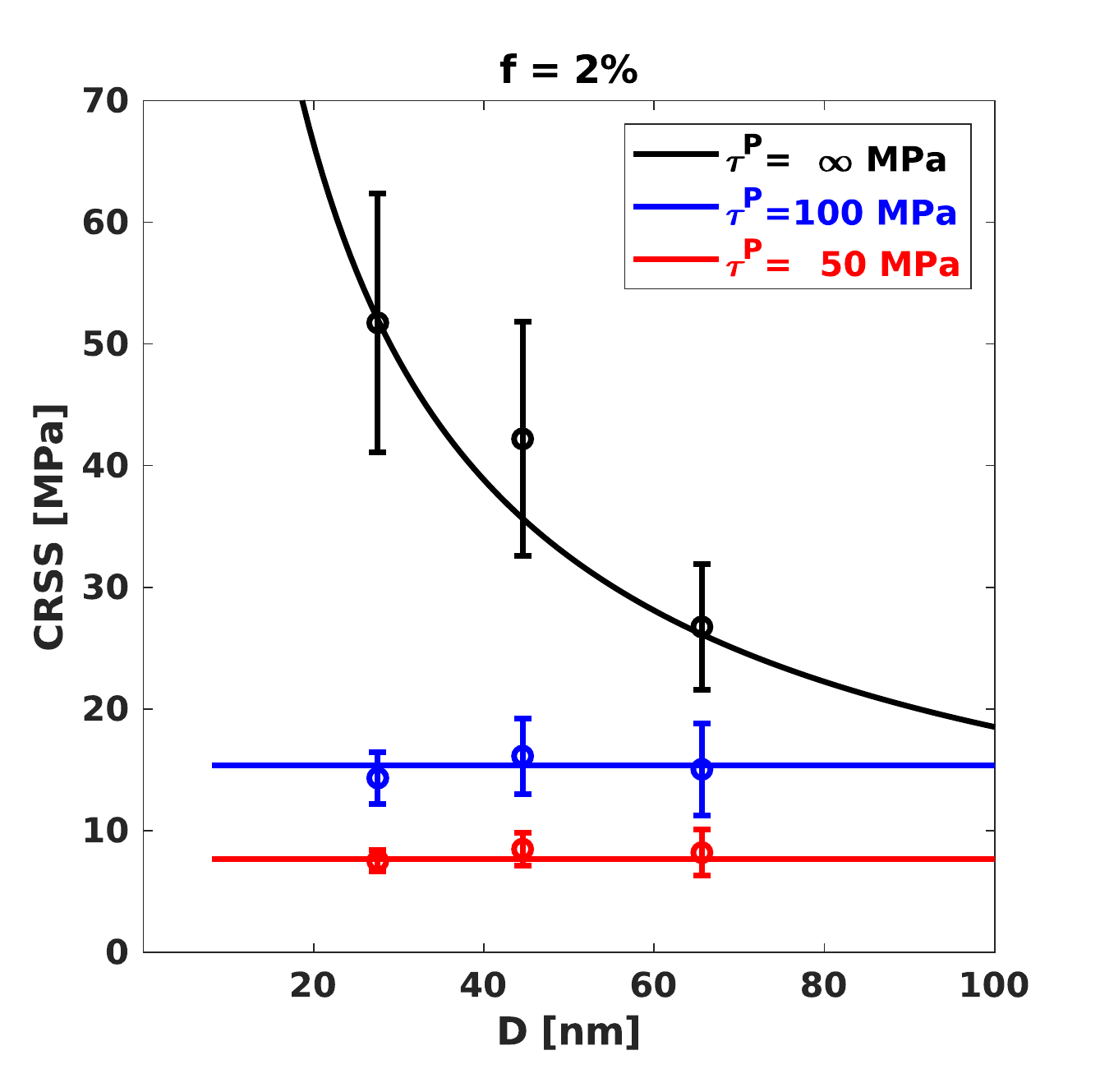}
	\caption{Comparison of the predictions of the CRSS with the generalized line tension tension model for a constant precipitate volume fraction ($f$ = 2\%)  with the results of DD simulations for random precipitate distributions. The average values from the DD simulations are plotted as open circles, along with the error bars indicating the standard deviation. The model predictions are included as solid lines for impenetrable (eq. 
\eqref{Eq:BKS_random_mean_heterogeneous}) and shearable precipitates (eq. \eqref{Eq:ShearableHeteroFric_Rand_volFrac})}
	\label{Fig:vol_frac_2pc}
\end{figure}

In order to confirm this result, 10 different random distributions of spherical precipitates with different diameters ($D$ = 27, 45 and 66 nm) and the same precipitate volume fraction (2\%) were generated in a domain with the same dimensions as above.  Obviously, the total number of precipitates in the distributions decreased as the precipitate diameter increased to keep constant the volume fraction. DD simulations were carried out using the same conditions indicated at the beginning of this section. Homogeneous precipitates, $G^P =G^M$, were considered in all cases. The average values and standard deviation of the CRSS obtained from the simulations are plotted as a function of the precipitate diameter $D$ in Fig. \ref{Fig:vol_frac_2pc}. They include the results of the simulations carried out with either an initial edge or screw dislocation. Therefore, the average values and standard deviations in Fig. \ref{Fig:vol_frac_2pc} were obtained from 20 DD simulations, 10 with an initial edge dislocation and 10 with an initial screw dislocation. One set of simulations was carried out assuming that the precipitates were impenetrable to dislocations and two other sets of simulations assumed shearable precipitates with $\tau^P$ = 50 MPa and 100 MPa. 

The predictions of eqs. \eqref{Eq:BKS_random_mean_heterogeneous} and \eqref{Eq:ShearableHeteroFric_Rand_volFrac} for random distributions of impenetrable and shearable spherical precipitates, respectively, are also plotted as solid lines in Fig. \ref{Fig:vol_frac_2pc}. The agreement between the generalized line tension model and the results of the DD simulations for shearable precipitates is excellent and confirms that the CRSS only depends on the precipitate volume fraction in this case. It should be noted that the predictions of the CRSS for shearable precipitates in eq. \eqref{Eq:ShearableHeteroFric_Rand_volFrac} are independent of the dislocation character when $G^P$ = $G^M$. In the case of heterogeneous shearable precipitates, the constants of the model (eq. \ref{Eq:ShearableHeteroFric_Rand_volFrac}) depend on the dislocation character, and a weighted average may be required.

It is worth noting that the dispersion of the CRSSs obtained from the DD simulations for different precipitate distributions mainly depends on $\tau^P$ (Figs. \ref{Fig:N12_edge}, \ref{Fig:N12_screw} and \ref{Fig:vol_frac_2pc}).  Weak precipitates are not significant obstacles to dislocation motion and they are overcome with small bow outs, leading to similar CRSSs regardless of the precipitate spatial distribution. On the contrary, impenetrable precipitates have to be overcome by the formation of dislocation loops that modify the shape of the dislocation line and, thus, the actual precipitate distribution determines up to a large extent the CRSS.

Finally, it should be noted that eqs. \eqref{Eq:BKS_random_mean_heterogeneous} and \eqref{Eq:ShearableHeteroFric_Rand} have been developed for monodisperse spherical precipitate distributions. It is likely that eq.  \eqref{Eq:ShearableHeteroFric_Rand} can also provide accurate results in the case of shearable precipitates with different shape and/or spatial distribution in so far $\tau^P$ is reduced. However, this may not be the case for impenetrable precipitates and the extension of this strategy to random distributions of precipitates with other geometries (such as disks, plates or rods) will be the objective of further investigations.

\section{Concluding remarks}

A generalized line tension model has been developed to estimate the CRSS in precipitation hardened alloys.  The model is based in previous line tension models for regular arrays of either impenetrable or shearable spherical precipitates that were expanded to take into account the effect of the elastic mismatch between the matrix and the precipitates. The parameters of the generalized line tension model were calibrated from dislocation dynamics simulations that covered a wide range of precipitate diameters and spacing as well as of the elastic constant mismatch. It was found that the CRSS increased with the ratio of the precipitate/matrix shear moduli in the case of impenetrable precipitates while differences  between the shear modulus of the matrix and the precipitate always led to an increase in the CRSS for shearable precipitates, regardless of whether the precipitate was stiffer or more compliant than the matrix.

The generalized line tension model for regular arrays of spherical precipitates was extended to deal with random arrays of monodisperse spherical precipitates by changing the geometrical parameters of the model by the averaged ones corresponding to the random distributions. The model predictions were in good agreement with the CRSSs obtained from dislocation dynamics simulations of random spherical precipitates distributions for both impenetrable and shearable precipitates. In the latter case, it was demonstrated that the CRSS was only a function of the precipitate volume fraction and independent of other geometrical parameters. The generalized line tension opens the way to carry out fast and accurate predictions of precipitate strengthening in metallic alloys taking into account the geometrical factors that characterize the precipitate distribution as well as the properties of matrix and precipitates (elastic constants, friction stress to shear the precipitate). Further extensions of the model should be aimed at including random distribution of precipitates with other geometries (such as disks, plates or rods) as well as other mechanisms that influence dislocation/precipitate interactions, such as the coherency strains.

\section*{Acknowledgments}

This investigation was supported by the European Research Council under the European Union's Horizon 2020 research and innovation programme (Advanced Grant VIRMETAL, grant agreement No. 669141). RSG acknowledges the support from the Spanish Ministry of Education through the Fellowship FPU16/00770.



\end{document}